\documentclass[aps,floatfix,preprint]{revtex4}

\usepackage{amsmath}
\usepackage{graphicx}
\usepackage{color}
\usepackage{pslatex}
\usepackage{setspace}

\usepackage[normalem]{ulem}

\newcommand{\se}{$\vert s \rangle\vert \epsilon\rangle$ }
\begin{document} 

\title{On Quantum Entropy and Excess Entropy Production in a System-Environment Pure State}

\author{Phillip C. Lotshaw}\thanks{currently at Oak Ridge National Laboratory}
\email{lotshawpc@ornl.gov}
\author{Michael E. Kellman}
\email{kellman@uoregon.edu}
\affiliation{Institute of Theoretical Science, University of Oregon \\ Eugene, OR 97403, USA}

\date{\today}

\begin{abstract}

We explore a recently introduced quantum thermodynamic entropy $S^Q_{univ}$ of a pure state of a composite system-environment computational ``universe"  with a simple system $\mathcal{S}$ coupled to a constant temperature bath $\mathcal{E}$.  The principal focus is ``excess entropy production" in which the quantum entropy change is greater than expected from the classical entropy-free energy relationship.  We analyze this in terms of   quantum spreading of time dependent states, and its interplay with the idea of a microcanonical shell. The entropy takes a basis-dependent Shannon information definition.  We argue for the zero-order $\mathcal{SE}$ energy basis as the unique choice that gives classical thermodynamic relations in the limit of weak coupling and high density of states, including an exact division into system and environment components.   Entropy production takes place due to two kinds of processes.  The first is classical ``ergodization"    that fills the full density of states within the microcanonical shell. The second is excess entropy production  related to quantum spreading or ``quantum ergodization" of the wavepacket that effectively increases the width of the energy shell.  Lorentzian superpositions with finite microcanonical shell width lead to classical results as the limiting case, with no excess entropy.  We then consider a single $\mathcal{SE}$ zero-order initial state,  as the examplar  of extreme excess entropy production.    Systematic formal results are obtained for a unified treatment of excess entropy production for  time-dependent Lorentzian superpositions, and verified computationally.  It is speculated that the idea of free energy might be extended to a notion of ``available energy" corresponding to the excess entropy production.    A unified perspective on quantum thermodynamic entropy is thereby attained from the classical limit to extreme quantum conditions. 

\end{abstract}

\maketitle

\section{introduction}

This paper concerns     ``excess entropy production"  in simulations of a system-environment pure state, using a recently devised quantum thermodynamic entropy for a pure state.   These ideas arise out of  new developments in quantum thermodynamics.   Recent years have seen a renewal of interest in the foundations of quantum statistical mechanics, both for theoretical reasons and practical interests of new technologies.  A prominent line of research has investigated  thermodynamic behavior in pure state systems \cite{micro,deltasuniv,twobath,variabletame,polyadbath,Belgians2,Olshanni,Gemmer:2009,Rudolph,tasaki1998}.   Arguments based on the ``eigenstate thermalization hypothesis" \cite{Deutsch, RigolReview, Greiner2016, Deutsch1991, DeutschSupplemental, Porras, Leitner2015, Leitner2018, Rigol} and ``typicality" \cite{Goldstein2006, Goldstein2010, vNcommentary, Reimann2008, Popescu2009, Popescu2006, Reimann2016, Goldstein2015} stake a persuasive claim that thermalization is a property of entanglement in pure states of complex systems.  Nonetheless, we have called attention to  an apparent gap in the quantum thermodynamics of pure states.  The classical second law states that in spontaneous processes, the ``entropy of the universe" is always increasing  

\begin{equation} \Delta S_{univ} > 0 \label{secondlaw}     \end{equation}

\noindent with the following relation between free energy and entropy change:  

\begin{equation}   - \frac{1}{T}    \   \Delta F_{sys} =  \Delta S_{univ}.          \label{greatresult}  \end{equation} 

\noindent However, these relations seem to be missing  in quantum thermodynamics of a pure state, because the standard von Neumann quantum entropy is zero for such a state.  In Refs.~\cite{deltasuniv,micro} we defined a quantum entropy $S^Q_{univ}$ for a pure system-environment (SE) state, compared this with $\Delta F_{sys}$, and found that we could recover the microcanonical limit Eq.~\ref{greatresult} for equilibration and thermalization.   These results seem to justify calling  $S^Q_{univ}$  a thermodynamic entropy for a pure state. To a certain extent, our approach goes back to interest of von Neumann himself in new ideas of quantum entropy \cite{vNtrans, vNcommentary}.   Others have introduced ideas  somewhat related to ours of the entropy of a pure state both for analyzing pure state thermodynamics \cite{HanEntropy,PolkovnikovicEntropy} and characterizing the information content of pure states \cite{KakEntropy,informationentropystotland}.  

 The  focus of the present paper is {\it excess entropy production}:  thermodynamic entropy $\Delta S^x$ beyond what is implied in the classical relation Eq.~(\ref{greatresult}).  This is a quantum phenomenon that was noted in Refs.~\cite{deltasuniv,micro}, in the course of exploring the classical microcanonical limit in which Eq.~\ref{greatresult} holds. Excess entropy production can lead to distinctly non-classical processes.  We have shown \cite{twobath} that these can include surprising phenomena such as heat flow from cold to hot and asymmetric temperature equilibration in a tripartite total system.      In the present work, the goal is a deeper systematic account of this excess entropy production and its physical meaning. A unified understanding is sought of the quantitative behavior of $\Delta S^x$  between  extreme limits of zero (i.e.~classical) and maximal,  massive excess entropy.  Concepts of density of states, quantum spreading of states, and Boltzmann entropy come into play.   It should be possible to control excess entropy by ``tuning" the initial state.   We anticipate that this understanding will be useful for devising and analyzing highly unusual, nonclassical situations in quantum thermodynamics of complex systems.  
 
An outline of the paper is likely useful.  Section \ref{SE section} summarizes the model system $\mathcal{S}$ and environment $\mathcal{E}$.   Section \ref{entropysection} deals with the definition of the quantum thermodynamic entropy for a pure $\mathcal{SE}$ state and Section \ref{sum of sys and env}  concerns the formal division of the entropy into system and environment parts.    Section \ref{heuristic section}  emphasizes heuristically that excess entropy production takes place  almost entirely in the environment, with the excess occurring due to quantum spreading of the wavepacket.  This insight is then examined computationally and formally in the succeeding sections.  Lorentzian wave packets turn out to be ideally suited for this.    In  Section  \ref{Lorentzian state section} time-evolving Lorentzian states are constructed, and their dynamics simulated computationally.  Section \ref {master entropy section} examines analytic and computational ``master relations" for the quantum entropy  $S^Q_{univ}$    and the excess entropy   $\Delta S^x$     in the near-classical and extreme-quantum limits. Section \ref{summary} gives a summing-up and ends with some speculation about excess ``free" or ``available" energy associated with excess entropy production.    

\section{Model system and environment}\label{SE section}

The model system $\mathcal{S}$ and environment $\mathcal{E}$ are taken from Ref.~\cite{micro}.  The system has evenly spaced levels.  The environment is designed to mimic a standard heat bath with a constant temperature $T$. Small environments with time- and size-dependent temperatures are also of interest in other contexts \cite{variabletame}, but our main concern here is to understand quantum thermodynamics in the standard limit where the bath is large. To model a large bath, we use  a discrete approximation to a true continuum of environment states, with density of states

\begin{equation}\label{rho} \rho_\mathcal{E}(E_\mathcal{E}) = Ae^{E_\mathcal{E}/T} \end{equation}

\noindent with the Boltzmann constant $k_B=1$. We consider the temperature as the fixed value $T=6.22$ energy units throughout the evolution of all $\mathcal{SE}$ states we consider.  We use a system with three evenly spaced energy levels $\{\vert s \rangle\} = \{\vert 0\rangle, \vert 1\rangle, \vert 2\rangle\}$ with energies $E_s = s$. The system and environment interact through a random matrix coupling with diagonal elements set to zero and off-diagonal elements taken as  random Gaussian variates scaled by a coupling constant $k$ that determines the strength of the interaction. Typically we take $A=1415.3$ and $k=0.9 \times 10^{-4}$, except in Sec.~\ref{micro limit section}. Details of the model are given in Ref.~\cite{micro}.

\begin{figure}
\begin{center}
\includegraphics[width=10cm,height=10cm,keepaspectratio,angle=0,trim={0cm 0cm 0cm 0cm},clip]{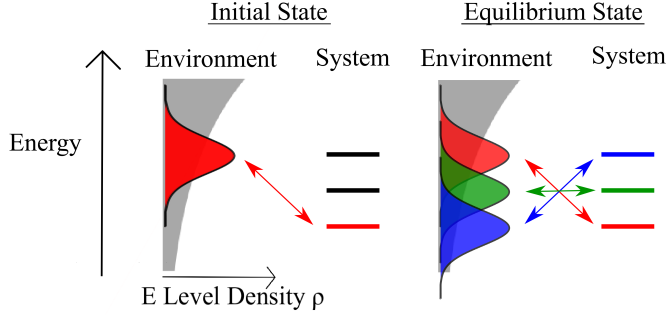}
\end{center}
\caption{Schematic of time evolution behavior, beginning with a superposition state in the environment and a single zero-order level in the system, then evolving to a thermalized set of entangled system-environment pairs.}
\label{schematic}
\end{figure}

A schematic of the time evolution  behavior is shown in Fig.~\ref{schematic}. The system begins in a single zero-order energy level $\vert s \rangle$, the environment begins in a superposition of zero-order levels $\vert \epsilon\rangle$ (we consider variations to environment state later, down to a single energy level $\vert \epsilon \rangle$), and the total density of $\mathcal{SE}$ states is $\rho_0 = \rho_\mathcal{E}(E_\mathcal{E})$. The system and environment then interact, exchange heat and evolve to equilibrium, resulting in the final state on the right.  In the final state there are a variety of entangled system-environment components, shown by matching colors in the diagram, each with approximately the same total energy $E = E_\mathcal{E} + E_\mathcal{S}$, with a total density of states $\rho_f(E) = \sum_s \rho_\mathcal{E}(E-E_s)$. The system alone is described by a thermal Boltzmann distribution $p_s \sim e^{-E_s/T}$, related to the exponentially increasing density of environment states at higher energy.  This is the schematic view of the system, environment, and thermalization process whose properties we examine in this paper.

\section{basis set and pure state quantum thermodynamic entropy}\label{entropysection}

First we consider the formal definition and rationale for the pure state quantum entropy $S^Q_{univ}$.    The definition of $S^Q_{univ}$ will depend on a choice of basis set. Our treatment of the basis set expands upon the more cursory justification offered in Ref.~\cite{deltasuniv}. Our choice is the zero-order $\mathcal{SE}$ energy basis.   In this and following sections we argue that this, and likely only this choice, will allow  recovery of classical microcanonical results, including the Boltzmann distribution, the canonical ensemble, the identity in Eq.~\ref{greatresult} in the limit of weak coupling, and the standard microcanonical relation $\Delta S_{env} = Q/T$ between entropy change of the environment and heat flow.   

To define the entropy  we first choose a ``reference basis"  $\{ | \alpha \rangle \}$.  In this basis a pure state is expressed as 

\begin{equation} | \Psi_\mathcal{SE} (t)  \rangle = \sum_\alpha c_\alpha (t) | \alpha \rangle.    \label{expansionagain}  \end{equation}  

\noindent   Then taking

\begin{equation}  p_\alpha (t) = | c_\alpha (t) |^2    \label{probability}    \end{equation}

\noindent we define the quantum entropy 

\begin{equation}  S^Q_{univ} = S^{\{\alpha\}} _{univ} = - \sum_\alpha p_\alpha \ln p_\alpha     \label{suniv}  \end{equation}    

\noindent with respect to the reference basis $ \{ | \alpha \rangle \}$.  This expression for the entropy has an evident relation to the Shannon information entropy.   In the quantum context it has been discussed as the ``conditional information entropy" by Stotland et al.  \cite{informationentropystotland}.    One of our goals has been to develop and exploit a thermodynamic meaning for this entropy.

This entropy depends on the choice of reference basis. We choose the  zero-order energy basis 

\begin{equation}    \{| \alpha \rangle \} = \{| s \rangle |\epsilon \rangle \}   \label{referencebasis}     \end{equation}  

\noindent of the $\mathcal{SE}$ complex.    Let the total state   be expanded in terms of zero-order system-environment energy bases   $\{\vert s\rangle\}$ and $\{\vert \epsilon \rangle\}$:  

\begin{equation} \vert \Psi\rangle = \sum_{s,\epsilon}c_{s,\epsilon}\vert s \rangle \vert \epsilon\rangle   \label{zoexpansion}      \end{equation}

\noindent with $\{\vert s\rangle\}$ and $\{\vert \epsilon \rangle\}$  the ZO basis sets of system and environment.   Eqs.~\ref{expansionagain}-\ref{zoexpansion} comprise the essence of our quantum entropy.  
What is the rationale for this?   We will be simulating a quantum total system $\mathcal{SE}$ in which energy flows between a system $\mathcal{S}$ and an environment $\mathcal{E}$ -- a process of heat flow. 
In statistical mechanics we typically have in mind  measurement of an $\mathcal{S}$ energy level, e.g.~ the energy  of a Brownian particle in a gravitational field.  This mirrors the choice of the zero-order system basis.   Then, if we are concerned with thermalizing energy flow,  the most natural further observation would be of the zero-order energy of $\mathcal{E}$ to give a total zero order energy of $\mathcal{SE}$.  This naturally leads to the basis of ZO energy states $\{| \alpha \rangle \} = \{| s \rangle |\epsilon \rangle \}$.  We might not be so interested in measuring the  energy of $\mathcal{E}$ -- that would be the usual case in the analogy to the Brownian particle.  Nonetheless, there  are further grounds to favor the $\mathcal{SE}$ zero-order basis.   We are naturally interested in constructs  that relate to the microcanonical ensemble for a fixed total energy.  Since we are interested in observing the $\mathcal{S}$ zero-order energy, the natural way of getting something like a total energy would seem to be as the sum E$_\mathcal{SE}$ = E$_\mathcal{S}$ + E$_\mathcal{E}$, i.e.~the sum of zero order energies, congruent with standard reasoning in the microcanonical ensemble.  Having singled out the $\mathcal{S}$ states and the energy sum $E$, the obvious basis is then the zero-order $\mathcal{SE}$ basis.  This justification seems compelling, but we will introduce further arguments based on the idea of a division of the entropy $S^Q$ into system and environment components, to which we turn in the next two sections. 

\section{$S^Q_{unv}$ as a sum of system and environment terms}  \label{sum of sys and env}

In this section we show how the quantum entropy of Eqs.~\ref{expansionagain}-\ref{zoexpansion} can be  divided into a sum of system and environment components. The system component comes from the reduced density matrix; the  environment component is an averaged sum of contributions,  weighted by system probabilities.  We begin with the general expression for the Shannon entropy of a bipartite system

\begin{equation} S = -\sum_{i,\lambda} p_{i,\lambda} \ln p_{i,\lambda}. \end{equation}

\noindent We will split this into separate parts for $i$ and $\lambda$, following Neilsen and Chuang \cite{nielsenchuangSenv}.  To begin, define the total probability for $i$ as

\begin{equation} \label{p i} p_i = \sum_\lambda p_{i,\lambda}. \end{equation}

\noindent Define the conditional probability for $\lambda$ when the first index is $i$ as

\begin{equation} \label{p lambda} p_{\lambda| i} = \frac{p_{i,\lambda}}{p_i} .\end{equation}

\noindent Using the newly defined probabilities and the normalization $\sum_\lambda p_{\lambda | i} = 1$ the entropy becomes \cite{nielsenchuangSenv}

$$ S = -\sum_{i,\lambda} p_ip_{\lambda | i} \ln p_ip_{\lambda | i} = -\sum_i p_i\ln p_i + \sum_i p_i\left(-\sum_\lambda p_{\lambda | i} \ln p_{\lambda | i}\right)  $$
\begin{equation} \label{Shannon decomposition} = S_i + \langle S_{\lambda }\rangle_{\{i\}}.\end{equation}

\noindent Eq.~\ref{Shannon decomposition} gives the general decomposition of the Shannon entropy of a bipartite system into separate parts for the two systems.  The entropy $S_i$ is a standard Shannon entropy for the first system.  The second system has a conditional entropy $\langle S_{\lambda }\rangle_{\{i\}}$ that is averaged with respect to the probabilities $p_i$ for the first system.  

Now consider the quantum entropy Eq.~\ref{suniv} with $\{\vert \alpha \rangle\} = \{\vert s \rangle\vert \epsilon\rangle\}$ as the reference basis of system-environment zero-order energy states and $p_\alpha = p_{s,\epsilon} = |c_{s,\epsilon}|^2$.  The entropy can be separated into system and environment parts, in parallel with Eq.~\ref{Shannon decomposition}:

\begin{equation} \label{Suniv decomposition} S^Q_{univ} = S^Q_\mathcal{S} +  S_\mathcal{E}^Q. \end{equation}

\noindent The system entropy

\begin{equation} S^Q_\mathcal{S} = -\sum_s p_s \ln p_s \end{equation}

\noindent uses system probabilities that can be calculated from the reduced density matrix with diagonal elements  $p_s = \sum_{\epsilon} p_{s,\epsilon} = \langle s \vert\hat \rho_\mathcal{S}\vert s \rangle$.  $S_\mathcal{S}^Q$ agrees with the standard quantum von Neumann entropy of the system when $\hat \rho_\mathcal{S}$ is dephased in the system zero-order energy basis $\{\vert s \rangle\}$, as in the Boltzmann thermal state and our initial state. The environment entropy is then

\begin{equation} \label{Senv} S_\mathcal{E}^Q = \sum_s p_s \left(-\sum_\epsilon p_{\epsilon | s} \ln p_{\epsilon | s}\right). \end{equation}

\noindent with $p_{\epsilon | s} = p_{s,\epsilon}/p_s$.  Eqs.~\ref{Suniv decomposition}-\ref{Senv} give our procedure  for calculating the total entropy $S^Q_{univ}$ in terms of system $S^Q_\mathcal{S}$ and environment $S_\mathcal{E}^Q $ parts.    

\section{entropy production in the environment:  density of states and quantum spreading } \label{heuristic section}

We have shown how to decompose the total entropy change $\Delta S_{univ}^Q$ into separate parts $\Delta S_\mathcal{S}^Q$ and $\Delta S_\mathcal{E}^Q$ for the system and environment.  Now we give a heuristic argument for how to relate these to their classical microcanonical counterparts $\Delta S_{sys}^\mathrm{micro}$ and $\Delta S_\mathcal{E}^\mathrm{micro}$.  We will find that excess entropy production is a component of the environment entropy change beyond the classical $Q/T.$   We will find later in Section \ref{master entropy section} that this correlates with analytical results for Lorentzian superposition states in our numerical simulations.

First consider the classical entropy change during system-environment thermalization.  The system and environment begin in isolation, corresponding to a microcanonical ensemble of $W_0 = \rho_0\delta E$ states with entropy $S_{univ,0}^\mathrm{micro} = \ln W_0$, where $\rho_0$ is the initial density of states and $\delta E$ is the width of the microcanonical energy shell.   The system and environment then exchange heat, evolving to fill a larger set of $W_f = \rho_f\delta E$ states.  The microcanonical entropy change is

\begin{equation} \label{delta suniv micro}  \ln \frac{\rho_f}{\rho_0} = \Delta S_{univ}^\mathrm{micro} = \Delta S_\mathcal{S}^\mathrm{micro} + \Delta S_\mathcal{E}^\mathrm{micro} \end{equation}

\noindent where the last equality uses Eq.~\ref{Shannon decomposition}.  The environment entropy change in Eq.~\ref{delta suniv micro} is given by the standard relation between the heat $Q$ and temperature $T$,

\begin{equation}\label{classical Q/T} \Delta S_\mathcal{E}^\mathrm{micro} = Q/T, \end{equation}

\noindent as shown in detail in Supplemental Information Sec.~A (Eq.~A22).  We now show how these quantities can be related to their quantum counterparts.

As anticipated in our previous work \cite{micro}, and developed analytically for time-dependent states in Section \ref{master entropy section} and Supplemental Information Sec.~C (Eq.~(C14)), the quantum entropy change can be analyzed in terms of a microcanonical-like relation   

\begin{equation} \label{SQmicro} S^Q_{univ} \sim \ln W_{eff} \end{equation}

\noindent with an effective number of states $W_{eff} = \rho\delta E$, and a variable effective energy width $\delta E$.  The width generally increases because of quantum state spreading during the dynamical equilibration process.  This results in a greater width for the final equilibrium state than the initial state $\delta E_f > \delta E_0$. Then the total entropy change is 

\begin{equation} \label{Delta Suniv Q micro} \Delta S_{univ}^Q \approx \ln\frac{\rho_f}{\rho_0} + \ln \frac{\delta E_f}{\delta E_0}        \end{equation}

\noindent     This is a key relation.  The  ``double logarithm  of ratios" form is very suggestive.  It is obtained heuristically here and analytically later in Eqs.~\ref{delta s Lorentzian},\ref{delta sx Lorentzian}   for time-evolving Lorentzian states. There are two sources of entropy change.  The first  term $\ln \rho_f / \rho_0$  is the classical system-environment entropy change from Eq.~\ref{delta suniv micro}.  It might be loosely referred to as due to ``classical ergodization."  The   second   term is the excess entropy production.  It is  due to quantum spreading or ``quantum ergodization"  of the energy shell:

\begin{equation} \label{sx spreading} \Delta S^x = \ln \frac{\delta E_f}{\delta E_0}.    \end{equation} 

We now use the system-environment decomposition of the entropy Eq.~\ref{Suniv decomposition} to show that $\Delta S^x$ is contained entirely within the environment.  Note that the quantum and classical entropy changes of the system are the same $\Delta S^Q_{sys} = \Delta S_{sys}^\mathrm{micro}$ since in both cases the system thermalizes to a Boltzmann distribution.  Then expressing $\Delta S_{univ}^Q$ on the left of Eq.~\ref{Delta Suniv Q micro} in terms of system and environment parts we have that  the quantum entropy change of the environment is

\begin{equation} \Delta S_\mathcal{E}^Q  \approx \Delta S_\mathcal{E}^\mathrm{micro}+  \ln \frac{\delta E_f}{\delta E_0} = \frac{Q}{T} + \Delta S^x \end{equation}

\noindent with the approximation indicated because this is a heuristic argument, in keeping with Eq.~\ref{Delta Suniv Q micro}.  The quantum entropy change of the environment is thus generally greater than the classical $Q/T$, with excess entropy production related to the increase in the width of the quantum energy shell.

With this analysis at hand, we return to the question of the justification for the $\mathcal{SE}$ zero order reference basis in defining $S^Q_{univ}$ in Eqs.~\ref{expansionagain}-\ref{referencebasis} of Section \ref{entropysection}.  Our decomposition of $S^Q_{univ}$ into system and environment parts gives standard results in the classical limit for a fixed microcanonical shell: $\Delta S_\mathcal{S}^Q = \Delta S_\mathcal{S}^\mathrm{micro}$ and $\Delta S_\mathcal{E}^Q = Q/T$.  Other choices of basis would give different values for the entropy changes.  This strongly supports the choice of the $\{\vert s \rangle\vert \epsilon\rangle\}$ reference basis as the unique basis that gives standard results in the classical microcanonical limit.

\section{Time evolving Lorentzian states}\label{Lorentzian state section}

Now we relate the preceding heuristic considerations to two concrete situations of Lorentzian wavepackets that are illuminating for being computationally transparent and analytically tractable.  We obtain analytical results for the entropy production from these states and confirm predictions in calculations, expanding on previous results \cite{micro} from a more heuristic approach.

Ref.~\cite{micro} observed in computations with different types of initial superposition states that there was excess entropy production.  In the weak coupling/infinite density of states limit this excess entropy went to zero, and classical microcanonical results were obtained: the free energy --  entropy relation Eq.~\ref{greatresult}.  Ref.~\cite{micro}  was about time evolution of a wave packet constructed from a superposition of many $\mathcal{SE}$ zero order states.  The initial wave packet had a single $\mathcal{S}$ state and many $\mathcal{E}$ states in the product.  The idea was to mimic an initial state reasonably close to classical,  within a microcanonical energy shell, with the width of the shell corresponding to the range of environment zero order energies.  The classical limit was obtained as the coupling $k$ and density of states $\rho$ were varied.  

However, it is not yet so clear under  what general conditions classical behavior will be recovered, and what governs the magnitude of excess entropy production.    What role is played by the width of the microcanonical shell, of both initial and final states, in determining the magnitude of excess entropy? Here we probe this by comparing the time evolution of two very different types of initial state, which are meant to serve as extreme cases.   One is a Lorentzian superposition of many $\mathcal{SE}$ initial states, corresponding to a microcanonical shell of significant width, with the expectation of quasi-classical behavior.      The second is a single zero order $\mathcal{SE}$ state.   Here  the width of the microcanonical shell is essentially zero.  The surmise  is that nonclassical effects of excess entropy production will be extreme with this latter  state.   

   \begin{figure*}
\begin{center}
\includegraphics[width=6.5cm,height=5.75cm,keepaspectratio,angle=0,trim={1.1cm 1cm 1cm 1cm},clip]{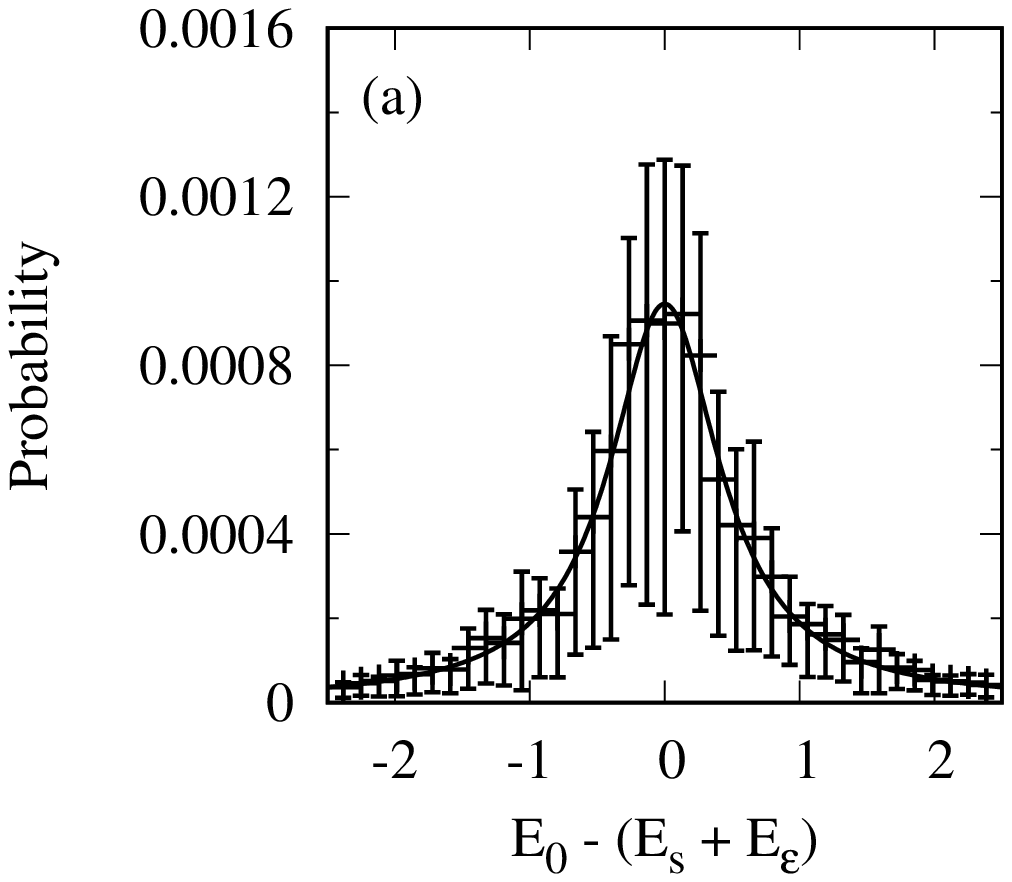}
\includegraphics[width=6.5cm,height=5.75cm,keepaspectratio,angle=0,trim={2.1cm 1cm 0cm 0.5cm},clip]{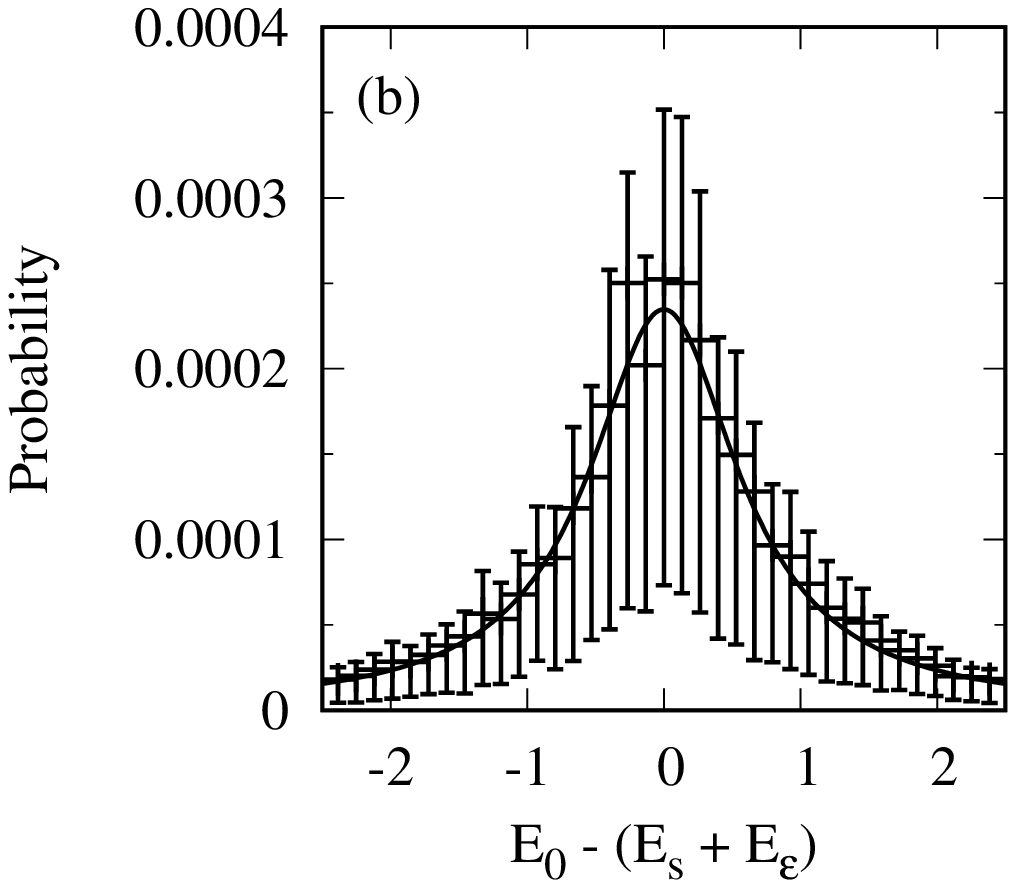}
\includegraphics[width=6.5cm,height=5.5cm,keepaspectratio,angle=0,trim={2cm 0 1cm 0},clip]{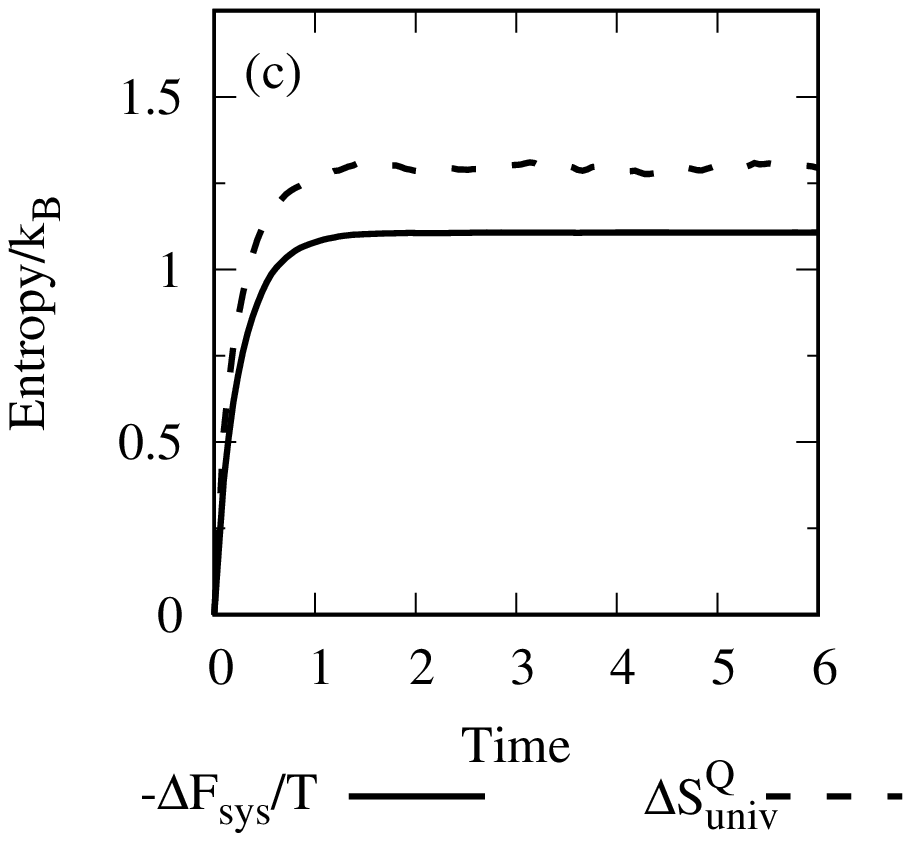}
\end{center}
\caption[Initial Lorentzian state, entropy production, and free energy change.]{Average probabilities $|\langle s \vert \langle \epsilon \vert \Psi \rangle|^2$ for nearby $\vert s \rangle \vert \epsilon\rangle$ basis states for (a) an initial Lorentzian state from Eq.~\ref{Lorentzian initial state} with half-width at half-max $\gamma_0 = 0.5$ and (b) the corresponding time-evolved state of Eq.~\ref{Lorentzian final state}.  The asymmetric error bars show the first and third quartiles of the distribution of probabilities that go into the averages shown by the data points.  (c) Entropy production $\Delta S_{univ}^Q$ and free energy change $-\Delta F_{sys}/T$ during the time evolution.}
\label{superdynamics}
\end{figure*}

   \begin{figure*}
\begin{center}
\includegraphics[width=6.5cm,height=5.75cm,keepaspectratio,angle=0,trim={1.1cm 1cm 1cm 1cm},clip]{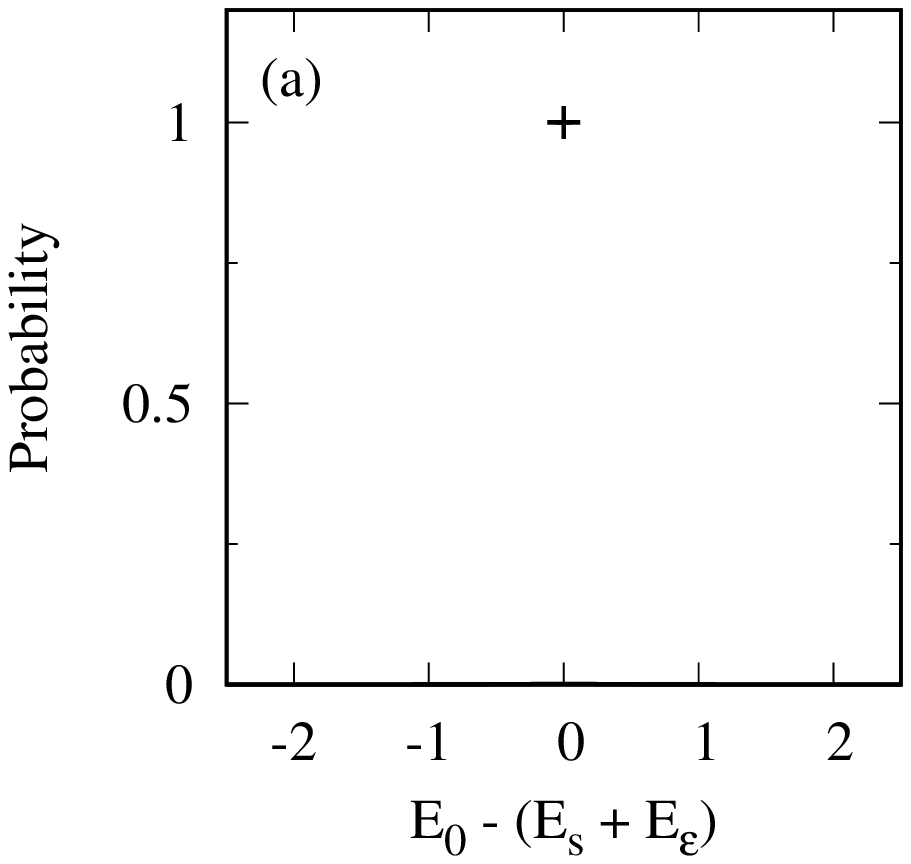}
\includegraphics[width=6.5cm,height=5.75cm,keepaspectratio,angle=0,trim={2.1cm 1cm 0cm 0.5cm},clip]{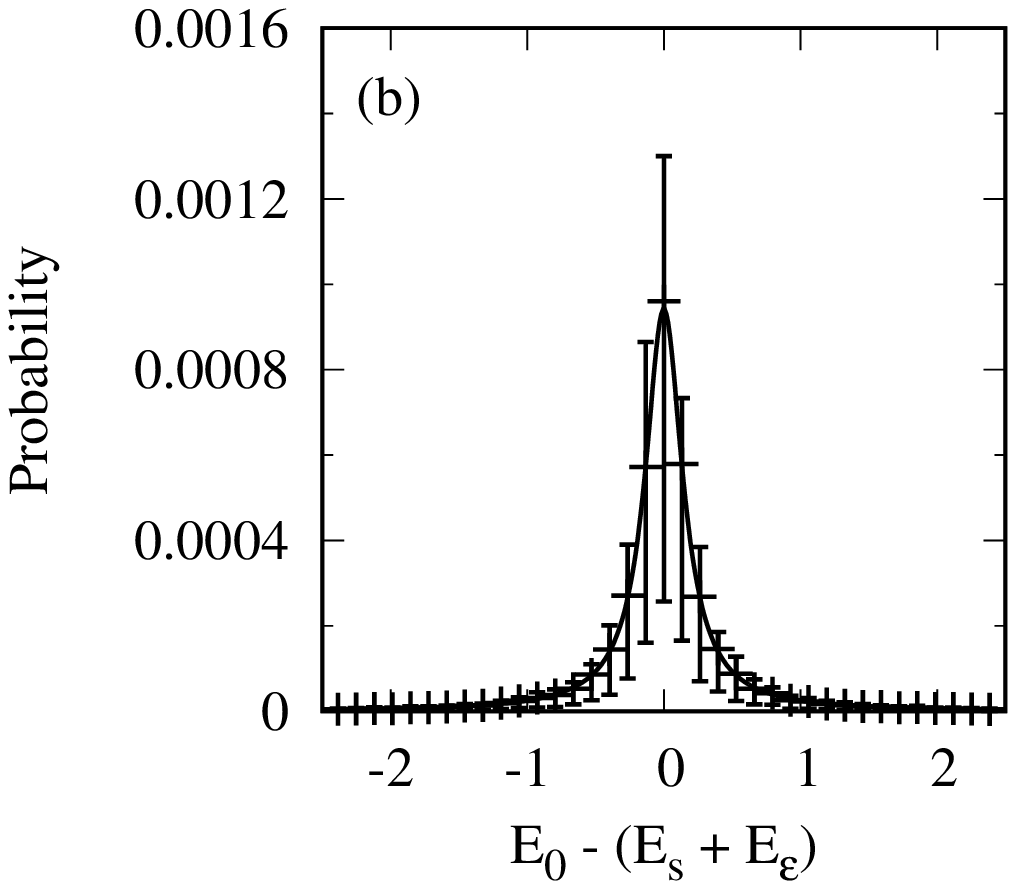}
\includegraphics[width=6.5cm,height=5.5cm,keepaspectratio,angle=0,trim={2cm 0 1cm 0},clip]{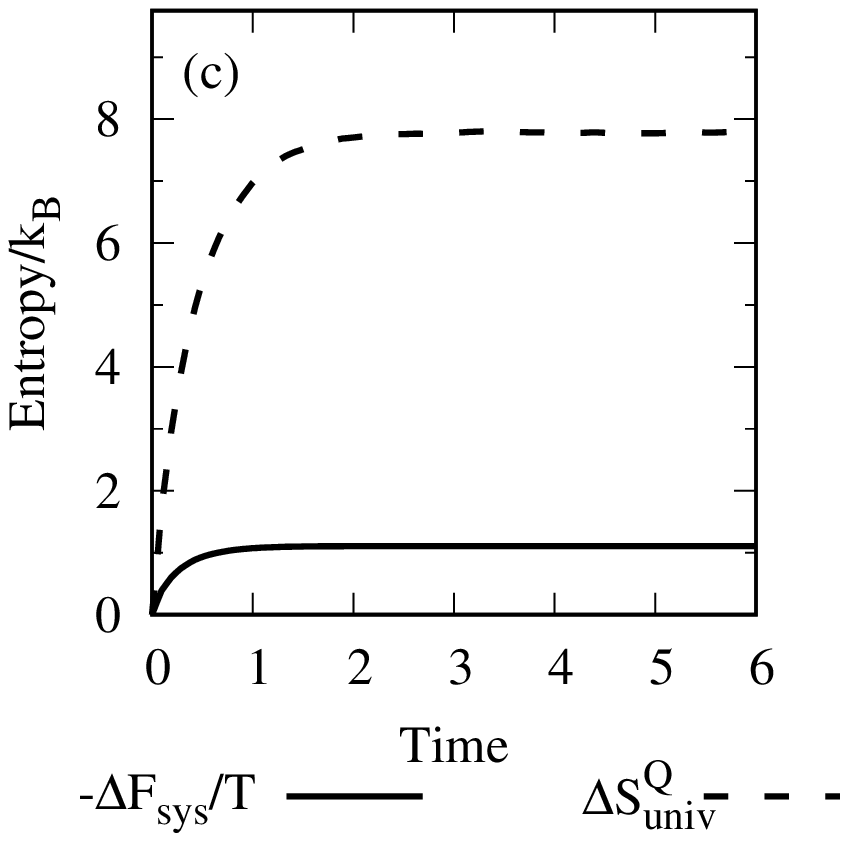}
\end{center}
\caption[Initial \se state, entropy production, and free energy change.]{Average probabilities $|\langle s \vert \langle \epsilon \vert \Psi \rangle|^2$ for nearby $\vert s \rangle \vert \epsilon\rangle$ basis states for (a) an initial $\vert s \rangle\vert \epsilon\rangle$ state from Eq.~\ref{se initial state} and (b) the corresponding time-evolved state of Eq.~\ref{Lorentzian final state}. The asymmetric error bars show the first and third quartiles of the distribution of probabilities that go into the averages shown by the data points.  (c) Entropy production $\Delta S_{univ}^Q$ and free energy change $-\Delta F_{sys}/T$ during the time evolution.}
\label{singledynamics}
\end{figure*}

\subsection{Initial states}

We investigate these questions with two kinds of simulations.   One takes a random superposition of $\mathcal{SE}$ zero-order states under an overall Lorentzian window.  The second takes a single $\mathcal{SE}$ zero-order state -- an extreme case of a Lorentzian, with zero width.  The results, shown in Figs.~\ref{superdynamics},\ref{singledynamics}, will be critically discussed in the following sections.  Reasons for considering Lorentzians are as follows.  The energy eigenstates are Lorentzian superpositions of zero order states, as pointed out by Deutsch in his introduction \cite{Deutsch1991} of the eigenstate thermalization hypothesis (ETH).  Following  analysis of Supplemental Information Sec.~B,  a single $\vert s \rangle\vert\epsilon\rangle$ zero-order state then consists of a Lorentzian superposition of eigenstates, which then evolves into a Lorentzian of zero-order states.   An  initial Lorentzian time-dependent superposition evolves into a (wider) Lorentzian superposition.  Thus, with Lorentzian initial states, down to the limit of zero-width Lorentzian $\vert s\rangle\vert \epsilon\rangle$ states, we evolve to Lorentzian final states.  This gives a unified class of states for systematic analysis. There are analytical results that can be brought to bear on the statistics of these Lorentzians, and also the entropy production.  Furthermore, the Lorentzian width has a nice correspondence to the idea of a microcanonical shell width.  Hence, Lorentzians are ideally suited for the systematic quantitative investigation of entropy production that we want to undertake.  

We will consider initial states with the system in a single level $\vert s \rangle$ and the environment described by fluctuations $\tilde g_{s,\epsilon}$ about a Lorentzian distribution $L^0_{s,\epsilon}$  

\begin{equation} \label{Lorentzian initial state} \vert \Psi^0_L\rangle \sim \sum_{\epsilon} \tilde g_{s,\epsilon} \sqrt{L_{s,\epsilon}^0} \vert s \rangle\vert \epsilon\rangle \end{equation}

\noindent The Lorentzian distribution at time $t=0$ is 

\begin{equation} \label{initial Lorentzian} L_{s,\epsilon}^0(E_s + E_\epsilon) = \frac{1}{\pi} \frac{\gamma_0/\rho_0}{(E_s + E_\epsilon -E_0)^2 + \gamma_0^2} \end{equation}

\noindent where $\rho_0$ is the density of environment states that pair with the initial system level $\vert s \rangle$ and $E_0$ and $\gamma_0$ are parameters that respectively describe the central energy of the Lorentzian and the half-width at half-max.  The $\tilde g_{s,\epsilon}$ are complex random gaussian variates that give random deviations to the $\vert s \rangle\vert \epsilon\rangle$ basis state probabilities about the Lorentzian average. These are taken as as  

\begin{equation} \label{gaussian variate} \tilde g_{s,\epsilon} = \frac{g_{s,\epsilon} + ig'_{s,\epsilon}}{\sqrt{2}} \end{equation}

\noindent where $g_{s,\epsilon}$ and $g'_{s,\epsilon}$ are random numbers from Gaussian distributions, e.g.~

\begin{equation} p(g_{s,\epsilon}) = \frac{1}{\sqrt{2\pi}} e^{-g_{s,\epsilon}^2/2}  \end{equation}

Panel (a) of Fig.~\ref{superdynamics} shows an example of an initial state with random variations about a Lorentzian as in Eq.~\ref{Lorentzian initial state} with an initial width $\gamma_0=0.5$.  The data points in the figure are averaged squared coefficients $|c_{s,\epsilon}|^2 = |\langle s\vert \epsilon\vert \Psi\rangle|^2$, averaged over all $\vert s \rangle \vert \epsilon\rangle$ states in an energy interval, with asymmetric error bars indicating the first and third quartiles of the coefficient distributions in the intervals. The average squared coefficients follow the Lorentzian from Eq.~\ref{initial Lorentzian}. The error bars are in good agreement with the quartiles expected from the Gaussian random deviations $\tilde g_{s,\epsilon}$, as discussed in detail in the Supplemental Information.

We will also be concerned with the time-evolution of initial single $\vert s \rangle \vert \epsilon\rangle$ basis states 

\begin{equation} \label{se initial state} \vert \Psi^0_{s,\epsilon}\rangle = \vert s \rangle \vert \epsilon \rangle \end{equation}

\noindent These can be viewed as the limit $\gamma_0 \to 0$ of the random Lorentzian initial states in Eq.~\ref{Lorentzian initial state}, where the Lorentzian distribution approaches a $\delta$ function.   Fig.~\ref{singledynamics}(a) shows an \se initial state as in Eq.~\ref{se initial state}, with just a single nonzero coefficient $c_{s,\epsilon}=1$.  This is an example of a very non-classical starting state, where the ``width" of the initial microcanonical energy shell is zero and the probability density is a delta function.

\subsection{Time evolution and quantum spreading}

Now consider the time evolution of the random Lorentzian states of Eq.~\ref{Lorentzian initial state} and of the single \se states of Eq.~\ref{se initial state}. We find in the Supplemental Information that both of these evolve to equilibrium states that can be described statistically by random fluctuations about a Lorentzian average, in accord with previous analyses of Deutsch \cite{Deutsch1991,DeutschSupplemental} and Nation and Porras \cite{Porras} in a similar model. The final states are given as

\begin{equation} \label{Lorentzian final state} \vert \Psi_L^{f}(t)\rangle, \vert \Psi_{s,\epsilon}^{f}(t) \rangle \sim \sum_{s,\epsilon} \tilde g_{s,\epsilon} \sqrt{L_{f}} \vert s \rangle\vert \epsilon\rangle \end{equation}

\noindent with a final state Lorentzian

\begin{equation} \label{final Lorentzian} L_f(E_s + E_\epsilon) = \frac{1}{\pi} \frac{\gamma_{f}/\rho_f}{(E_s + E_\epsilon -E_0)^2 + \gamma_f^2} \end{equation}

\noindent where $\rho_f$ is the total density of $\vert s \rangle\vert \epsilon\rangle$ states and $\gamma_f$ is the half-width at half-max of the final Lorentzian, to be discussed further below.

The half-width at half-max $\gamma_f$ is found in the Supplemental Information to be increased by the ``spreading factor" $2\pi k^2 \rho$ relative to the initial state width $\gamma_0$:   

\begin{equation} \label{gamma final} \gamma_f = \gamma_0 + 2\pi k^2 \rho_f, \end{equation}

\noindent where $k$ is the coupling strength and $\rho_f(E) = \sum_s \rho_\mathcal{E}(E-E_s)$ is the total density of states at equilibrium, following Sec.~\ref{SE section}.  Eqs.~\ref{Lorentzian final state}-\ref{gamma final} apply to both the equilibrated, time-evolved Lorentzian initial states of Eq.~\ref{Lorentzian initial state} and the time-evolved $\vert s \rangle \vert \epsilon\rangle$ states of Eq.~\ref{se initial state}, where for the latter it is understood that the  value $\gamma_0=0$ is used in the final width in Eq.~\ref{gamma final}, so that $\gamma_f = 2\pi k^2 \rho_f$. 

Panel (b) of Figs.~\ref{superdynamics} and \ref{singledynamics} show the time-evolved states.   Both states spread in time.  Both evolve to random fluctuations about the Lorentzians $L_f$ from Eq.~\ref{final Lorentzian} with the appropriate widths $\gamma_f$ from Eq.~\ref{gamma final}.  The variations in the coefficients are very well characterized by the Gaussian random variations $\tilde g_{s,\epsilon}$ in Eq.~\ref{Lorentzian final state}.  See the Supplemental Information Section~B for details.

\subsection{Entropy Production for the Time-Evolving \se and Lorentzian States}

Now we consider the entropy production for these examples of time-evolving states.  Panel (c) of Fig.~\ref{superdynamics} shows the entropy production  $\Delta S_{univ}^Q$ as the initial Lorentzian superposition in (a) evolves to the wider final Lorentzian distribution in (b).   $\Delta S_{univ}^Q$ is compared with the classical entropy change $-\Delta F_{sys}/T = \Delta S_{univ}^\mathrm{micro}$, which we compute using the standard definition $F_{sys} = \langle E_\mathcal{S}\rangle - TS^{vN}_\mathcal{S}$, with $S^{vN}_\mathcal{S}$ the von Neumann entropy.  There is some excess entropy production, as expected from Ref.~\cite{micro}, but overall it is fairly close to microcanonical. Figure \ref{singledynamics}(c) shows the entropy production for the initial single \se state that evolves to a final random Lorentzian.  Now there is a very large amount of excess entropy production.  It seems  that the finite microcanonical shell width of the initial state in Fig.~\ref{superdynamics} plays an essential role in getting the approach to classical behavior, because it limits the {\it relative} spreading of the wave packet in time, as suggested by Eq.~\ref{sx spreading}. To understand these connections systematically, we will take advantage of analytic expressions  for superposition states with a Lorentzian profile, using results from Supplemental Information Sec.~B. We will see that considerable insight is gained following this path. We will see that we can ``tune" the excess entropy production between the classical limit of zero for a suitable Lorentzian, and the very large excess of a single \se state, as suggested in Figs.~\ref{superdynamics} and \ref{singledynamics} and seen systematically in Figs.~\ref {suniv Lorentzian fig} - \ref{sx micro limit fig}, to which we turn next.

\section{master relationships for $S^Q_{univ}$ and $\Delta S^x$ for time-evolving Lorentzian states} \label{master entropy section}

Now we want to attain a  systematic understanding of the entropies in    these  simulations and how they change during equilibration, in comparison with the analytical results for the initial and final state statistics in Section \ref{Lorentzian state section}.     The basic results, seen in Figs.~\ref {suniv Lorentzian fig} - \ref{sx micro limit fig}, show evident regularities, which we briefly remark upon before detailed consideration.  The initial states are either random Lorentzians or a single \se basis state, as in Eqs.~\ref{Lorentzian initial state} and \ref{se initial state}.  We will find that the entropies for these two types of initial states can be united in the   analytic  ``master entropy" Eq.~\ref{suniv master}, which accounts for the pattern of Fig.~\ref {suniv Lorentzian fig}.  Both types of initial states evolve to final states that are also random Lorentzians, as in Eq.~\ref{Lorentzian final state}.  This will lead to an analytical expression for the expected excess entropy production in Eq.~\ref{sx master}  and Fig.~\ref{sx Lorentzian fig}. Finally, we consider the question of entropy production in approach to the classical limit, building on our earlier investigation of related but distinct questions in Ref.~\cite{micro}.  Fig.~\ref{sx micro limit fig} will show that superposition states approach classical entropy production $\Delta S^x = 0$ while the \se initial states are at a much different,  opposite extreme.  We will be  able to account for the regularities in Figs.~\ref {suniv Lorentzian fig} - \ref{sx micro limit fig} in accord with our earlier heuristic arguments about $S^Q$ and   $\Delta S^x$ in Section \ref{heuristic section}, based on ideas about the microcanonical shell and quantum spreading of the environment state during equilibration.  We begin with analysis of entropy production and excess entropy production with a fixed model for the environment and $\mathcal{SE}$ coupling in Figs.~\ref{suniv Lorentzian fig}-\ref{sx Lorentzian fig}, then analyze how variations in the size and coupling strength affect the entropy production for Lorentzian and \se initial states in Fig.~\ref{sx micro limit fig}.

\subsection{Entropy of the States}

\begin{figure}
\begin{center}
\includegraphics[width=7.5cm,height=7.5cm,keepaspectratio,angle=0]{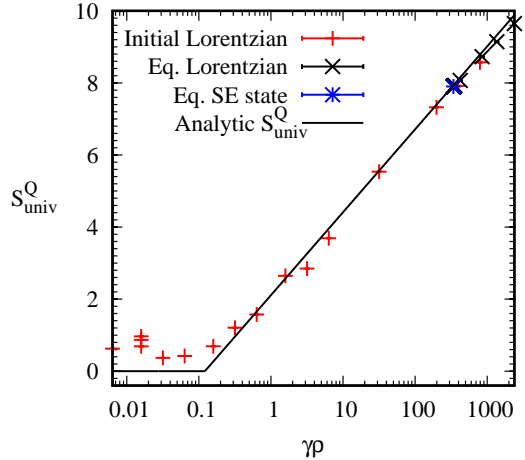}
\end{center}
\caption{Entropies from initial Lorentzian states (Eq.~\ref{Lorentzian initial state}) with varying widths $\gamma_0$, time-evolved version of these states at equilibrium, and a time-evolved \se state at equilibrium all approximately follow the analytic $S^Q_{univ}$ from Eq.~\ref{suniv master}.} 
\label{suniv Lorentzian fig}   
\end{figure}

We will find an analytic  relationship for the entropy of the  Lorentzian states, related to Boltzmann's entropy formula $S = k \ln W$ and the idea of a microcanonical shell width.   This is attained following the derivation in Supplemental Information Sec.~C (Eq.~(C.14)). The results, displayed  in  Fig.~\ref{suniv Lorentzian fig},  show an interesting relationship between the analytic predictions and computed entropies.   We now outline the derivation and meaning of these results.  The derivation approximates the fundamental sum for the entropy Eq.~\ref{suniv} as an integral over the random Lorentzian coefficients, such as appear in Eqs.~\ref{Lorentzian initial state} and \ref{Lorentzian final state}.  The integral approximation is justified when the Lorentzian is wide relative to the $\mathcal{SE}$ energy level spacing.  The result is that the entropy for the Lorentzian is 

\begin{equation} \label{s Lorentzian} S_{L} = \ln (4 \pi \gamma \rho) - g_0 \end{equation}

\noindent where $\rho$ is the density of states and $\gamma$ is the half-width at half-max of the Lorentzian.  These have the values $\rho_0$ and $\gamma_0$ for the initial Lorentzian states, as appear in Eq.~\ref{initial Lorentzian},  and  $\rho_f$ and $\gamma_f$ for the final states, as described in the discussion around Eq.~\ref{gamma final}.   

The first term on the right of Eq.~\ref{s Lorentzian} gives the entropy of a perfect Lorentzian (without the random variations $\tilde g_{s,\epsilon}$).  This has the microcanonical-like form suggested previously in Eq.~\ref{SQmicro}, 

\begin{equation} \ln (4\pi \gamma\rho) = \ln (\rho \delta E ) = \ln W_{eff} \end{equation}

\noindent with $W_{eff} = \rho\delta E$ an effective number of states in an energy shell of width $\delta E = 4\pi \gamma$. We compute the second term in (\ref{s Lorentzian})   using {\it Mathemtica} and find

\begin{equation} g_0 = \langle |\tilde g_{s,\epsilon} |^2 \ln  |\tilde g_{s,\epsilon} |^2 \rangle = 1 - \gamma_{EM}. \end{equation}

\noindent This gives the deviation from the Lorentzian entropy due to the random fluctuations in the basis state probabilities $\tilde g_{s,\epsilon}$, with $\gamma_{EM} = 0.577\ 215...$ the Euler-Mascheroni constant \cite{WolframEulerMascheroni} (See Supplemental Information Sec.~C for details).  Using this value of $\gamma_{EM}$, $g_0 = 0.422 \ 785 ...$. We have thus obtained the desired relationship between $S^Q_{univ}$, Boltzmann's entropy formula, and the number of states with a given shell width and density of states.

How well this works is seen numerically  in Fig.~\ref{suniv Lorentzian fig} which shows the computational entropies for three kinds of initial states:  a time-evolved $\vert s \rangle \vert \epsilon\rangle$ state;  initial states with random variations about a Lorentzian, from Eq.~\ref{Lorentzian initial state}, with various initial widths $5 \times 10^{-6} \leq \gamma_0 \leq 0.25$;  and the time-evolved Lorentzian states with final widths $\gamma = \gamma_f$ from Eq.~\ref{gamma final}. The simulation results are well described by the approximate $S_L$ of Eq.~\ref{s Lorentzian} along the diagonal line of the figure, when $\gamma\rho$ is not too small.  For small $\gamma\rho$ on the left of the figure, the initial states  approach the limit of the single \se basis state, which of course has entropy $\ln(1) = 0$.   In this limit, the integral approximation that goes into the derivation of $S_L$ breaks down.  It gives $S_L <0$ when $\gamma < e^{g_0}/4\pi\rho$, whereas $S^Q_{univ} \geq 0$ by definition.  A better approximate formula is obtained by setting the entropy to zero when $S_L$ becomes negative, essentially approximating the entropy of the very narrow states by the value 0 for a single \se state.    This gives the complete generic ``master" formula for the entropy of the time evolving random Lorentzian states, plotted as a solid line:

\begin{equation}  \label{suniv master} S^{Q}_{univ} \approx \left\{ 
\begin{array}{lll}
\ln(4\pi \gamma\rho) -g_0 & : & \gamma \geq e^{g_0}/4\pi\rho \\
0 & : & \mathrm{otherwise}\\
\end{array} \right.
  \end{equation}

\noindent This is the same as the previous relation $S_L$, except that it stops changing when it reaches the minimum value zero, giving the abrupt bend in the figure.  The simulation results are in good agreement with this predicted behavior, with fluctuations around $S^Q_{univ} = 0$ at small $\gamma\rho$, and following the curve for $S_L$ at larger $\gamma\rho$.  The results show the master relation  for $S^Q_{univ}$ holds for initial Lorentzian states and time-evolved versions of those states.  For an initial \se state the entropy is zero in accord with the master relation, and the figure shows that a time evolved \se state is also in agreement with the general trend.  In sum, the approximate master entropy relation Eq.~\ref{suniv master} is giving a good account of  the simulation results for all the initial and final states we consider. We next use this relation to understand the excess component $\Delta S^x$ of the entropy production, as depicted in Fig.~\ref{sx Lorentzian fig}.   Finally, we analyze this entropy production in the question of   approach to classical microcanonical behavior  in the standard limit of a large bath and weak coupling, finding diverging results for ``normal" superpositions and ``extreme"  $\vert s \rangle \vert \epsilon\rangle$ initial states in Fig.~\ref{sx micro limit fig}.  

\subsection{Entropy Production}\label{sx results section suniv}

We next consider entropy production, including excess entropy production, during the approach to equilibrium. The aim is to see how entropy production relates to the intuitive ideas of the width of the microcanonical shell and the  of spreading of the quantum wave packet.    First, we consider initial states with random variations about a Lorentzian from Eq.~\ref{initial Lorentzian} that are well described by the approximate entropy $S_L$ of Eq.~\ref{s Lorentzian}, when $\gamma_0 \geq e^{g_0}/4\pi\rho_0$ in Eq.~\ref{suniv master}. The entropy change in time evolution from initial to final state is

\begin{equation} \Delta S_L = \ln(4\pi \gamma_f \rho_f) - g_0 -\left(\ln (4\pi \gamma_0 \rho_0) - g_0\right) \end{equation}

\noindent based on the initial and final values of $S_L$ from Eq.~\ref{s Lorentzian}. Cancelling terms gives

\begin{equation} \label{delta s Lorentzian} \Delta S_L = \ln \frac{\rho_f}{\rho_0} + \ln\frac{\gamma_f}{\gamma_0}. \end{equation}

\noindent This is the second appearance of the ``double logarithm of ratios" form,  noted earlier in connection with Eq.~\ref{Delta Suniv Q micro}.   The first term $\ln \rho_f/\rho_0$ gives the classical entropy change from heat flow, following the microcanonical definition Eq.~\ref{delta suniv micro}, in a process of ``classical ergodization."   The second term gives the quantum excess entropy production 

\begin{equation} \label{delta sx Lorentzian} \Delta S_L^x = \ln \frac{\gamma_f}{\gamma_0} = \ln \frac{\delta E_f}{\delta E_0} \end{equation}

\noindent due to quantum spreading or ``quantum ergodization"  of the environment state wave packet.  This analytic relation is similar to the somewhat more complicated empirical curve for fitting $\Delta S^x$ in Ref.~\cite{micro} with a less structured type of $\mathcal{SE}$ state, which did not maintain a consistent Lorentzian profile as we have here.  For our Lorentzians we obtain the simple formula of Eq.~\ref{delta sx Lorentzian}, in terms of only the initial and final Lorentzian widths $\gamma_0$ and $\gamma_f$. This corresponds transparently  to the increase in the effective width of the energy shell, as anticipated in Eq.~\ref{sx spreading}.

\begin{figure}
\begin{center}
\includegraphics[width=7.5cm,height=7.5cm,keepaspectratio,angle=0]{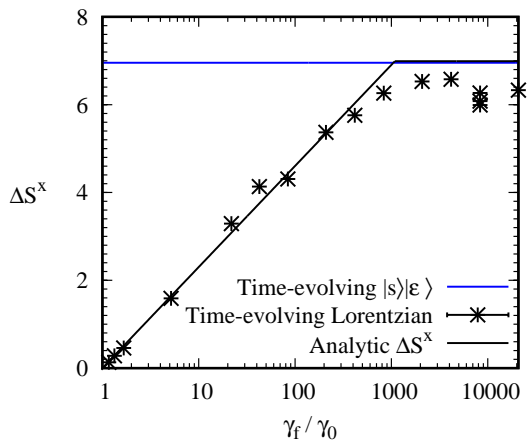}
\end{center}
\caption{Excess entropy production for initial random Lorentzians (black stars) and for an initial $\vert s \rangle\vert\epsilon\rangle$ state (blue bar) follow the master equation Eq.~\ref{sx master}.}
\label{sx Lorentzian fig}
\end{figure}

The diagonal line on the left of Fig.~\ref{sx Lorentzian fig} shows the analytic   $\Delta S_L^x$ of Eq.~\ref{delta sx Lorentzian} compared with $\Delta S^x = \Delta S_{univ}^Q + \Delta F_{sys}/T$ from the simulations.  Moving from left to right in the figure, we are decreasing $\gamma_0$ to increase the ratio $\gamma_f/\gamma_0$. The approximate relation is giving a  good account of the results on the left of the figure, where $\gamma_f/\gamma_0$ is not too large.

Now consider the right side of Fig.~\ref{sx Lorentzian fig}.  The $\Delta S^x$ are reaching close to the  maximum value, i.e.~ for a single \se state, corresponding to the limit of small $\gamma_0$ with large $\gamma_f/\gamma_0$.  For small $\gamma_0$, we want to take the initial state as the limiting single \se state, like what we did for $S^Q_{univ}$ in Eq.~\ref{suniv master}.   For a single \se initial state the initial entropy is zero.  The final state has the Lorentzian entropy $S_L$.  Then $\Delta S_{univ}^Q = S_L$. The maximum excess entropy production is then calculated by subtracting the microcanonical $\ln \rho_f/\rho_0$ ,

\begin{equation}  \Delta S^{x,max} = \ln (4\pi\gamma_f\rho_f) - g_0 - \ln \frac{\rho_f}{\rho_0}. \end{equation}

\noindent To evaluate this equation, we use $\gamma_f$ from Eq.~\ref{gamma final}, with $\gamma_0 = 0$ for the \se initial state.  This gives 

\begin{equation} \label{sx max} \Delta S^{x,max} =  \ln (8\pi^2k^2\rho_f\rho_0) - g_0. \end{equation}

\noindent We now have the ``master" equation for the excess entropy production

\begin{equation}  \label{sx master} \Delta S^{x} \approx \left\{ 
\begin{array}{lll}
\ln (\gamma_f/\gamma_0) & : & \gamma_0 \geq e^{g_0}/4\pi\rho_0 \\
\ln (8\pi^2k^2\rho_f\rho_0) - g_0 & : & \mathrm{otherwise}\\
\end{array} \right.
  \end{equation}
  
\noindent This master relation for $\Delta S^x$ is shown by the black solid line in Fig.~\ref{sx Lorentzian fig}.  It follows $\Delta S^x_L$ from Eq.~\ref{delta sx Lorentzian} up until this reaches the maximum value for a single \se initial state, where the master relation bends and becomes flat in the right of the figure.  This is in good agreement with our simulation results, which follow $\Delta S^x_L$ in the left of the figure then fluctuate around the maximum value in the right of the figure. In sum, we are getting a good systematic account of the excess entropy production.

\subsection{Excess Entropy Production and the Microcanonical Limit}\label{micro limit section}

\begin{figure}
\begin{center}
\includegraphics[width=7.5cm,height=7.5cm,keepaspectratio,angle=0]{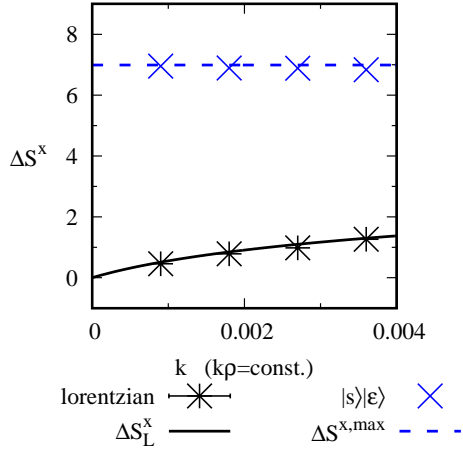}
\end{center}
\caption{Excess entropy production for Lorentzian initial states with widths $\gamma_0 = 0.0625$ and \se initial states, with  model environments having variable $k \rho$ constant, heading towards the microcanonical limit $k\to0, \rho\to\infty, k\rho = const.$  in the bottom left.  The Lorentzians approach the classical result $\Delta S^x = 0$, the \se states do not.}
\label{sx micro limit fig}
\end{figure}

We have seen that the source of  excess entropy production is the relative increase in the width of the environment state from quantum spreading during equilibration.  Larger  $\gamma_f/\gamma_0$ gives greater deviations from a fixed microcanonical energy shell, with larger $\Delta S^x$.  What has not been brought out so far is the role of the size of the environment and the $\mathcal{SE}$ coupling strength.  We have been dealing with a finite model environment with finite coupling, in contrast to the textbook situation with an infinite environment $\rho_f\to\infty$ and negligible coupling $k\to0$. Ref.~\cite{micro} showed with superposition states that microcanonical results $\Delta S^x =0$ were obtained in the  limit $k\to 0, \rho_f \to \infty$  with $k\rho_f = const$,  as needed to maintain thermalization within the simulations.  Ref.~ \ref{sx micro limit fig} had a quite different context than the analysis of Figs.~\ref{suniv Lorentzian fig} and \ref{sx Lorentzian fig},  which have a fixed density of states and coupling, and the extreme range of initial width  from wide superpositions to single \se initial states.  Now the question is how the choice fo the initial shell width affects the approach the classical limit $\Delta S^x =0$ as we increase the bath size and decrease the coupling. 

 Fig.~\ref{sx micro limit fig} shows $\Delta S^x$ in a series of calculations heading toward the supposed ``microcanonical limit" $k\to 0, \rho_f \to \infty$, with $k\rho_f = const$; this is accomplished by taking the baseline values for $k$ and the density of states prefactor $A$, noted below Eq.~(\ref{rho}), and varying these with $A \sim 1/k$ and $Ak=const.$. First consider the Lorentzian states in the figure, with small values of $\Delta S^x$.  These clearly approach classical behavior $\Delta S^x = 0$ from the right  as quantum spreading becomes negligible, with $\gamma_f \to \gamma_0$ in Eq.~\ref{gamma final} and therefore $\Delta S^x \to 0$ in Eq.~\ref{delta sx Lorentzian}.  Now consider the \se states in the figure.  They have very nearly constant computational $\Delta S^x \approx  \Delta S^{x,max} $, as predicted by Eqs.~\ref{sx max} and \ref{sx master}.  The maximum $\Delta S^x$ for an \se initial state depends only on the products $k\rho_f$ and $k \rho_0$, which are both constant as we approach the limit, so there is no hint at all of approach to classical behavior.  Instead, quantum spreading is always significant and  classical behavior is never observed for the \se states. For intermediate cases between the two extremes in Fig.~\ref{sx micro limit fig}, it should be possible to ``tune" the excess entropy, as emphasized  previously.

This completes our investigation of entropy content of states, entropy production including excess, and the question of what kinds of states approach the classical limit, within our range of ``normal" and ``extreme" states.  

\section{summary and concluding remarks}  \label{summary}

We have systematically explored the phenomenon of excess entropy production in time-dependent equilibration processes, in terms of the quantum thermodynamic entropy of Refs.~\cite{micro,deltasuniv} for a system-environment pure state.  Our focus is the role of quantum spreading of the $\mathcal{SE}$ state, its relation to the width of the microcanonical energy shell, and violation of  the classical entropy-free energy relation in the form of excess entropy production.  We span the range from near-classical behavior to the extreme of excess entropy with a single zero-order initial state.  

Using the Shannon information entropy definition, we defined the quantum entropy $S^Q_{univ}$  in terms of the zero-order energy basis.  This choice of basis is made on the grounds that thermodynamically one would be interested in observation of the system zero-order state, so  that straightforward definition of the energy in the microcanonical shell  necessitates the sum of zero-order system and environment energies. We showed that there is a mathematical division of $S^Q_{univ}$ into system and environment terms. With this, we found that our choice of basis for the definition of $S^Q_{univ}$ uniquely gives standard thermodynamic results in the classical limit of weak coupling and large density of states, with the standard classical relations $\Delta S_{univ} = - \Delta F_{sys}/T$  and $\Delta S_{env} = Q/T$.  The entropy is readily understood with Boltzmann's equation $S = k \ln W$, with $W$ being given by shell width $\times$ density of states $\delta E \times \rho = 4 \pi \gamma \rho$   according to Eq.~ \ref{s Lorentzian}.    

$\Delta S^Q_{univ}$ can be understood for Lorentzian states, as being due to two components.  One is ``classical ergodization" as the system thermalizes and heat flows into the environment, with consequent increase in the occupied density of states, giving the contribution  $\ln \rho_f/\rho_0$.  The second is excess entropy production $\Delta S^x$  due to quantum spreading or ``quantum ergodization"  of the microcanonical shell, represented by $\ln \gamma_f/\gamma_0$  in Eq.~ \ref{sx master}.  The excess happens in the environment -- not as heat flow $Q/T$ within the original,  ``classical" microcanonical shell, but rather as quantum spreading of the  shell. The final width of the Lorentzian $\gamma_f$ increases by an additive factor independent of the initial width $\gamma_0$, so the initial width  $\gamma_0$  is a critical factor in the ratio $\ln \gamma_f/\gamma_0$ that determines $\Delta S^x$. Hence, initial states like a microcanonical wave packet with  small relative spreading $\gamma_f/\gamma_0 \approx 1$  approach the classical limit.  On the other hand, initial states that approach the extreme limit of a single $\mathcal{SE}$ zero order state have maximal, massive entropy production, very different from classical. 

In sum, we have the following  picture.  The quantum entropy $S^Q_{univ}$ describes time-dependent thermodynamic evolution.    In the limit of small coupling, large bath, and non-zero initial energy-shell width, the classical limit is recovered.  Away from this limit, there is excess entropy production $\Delta S^x>0$.  This excess is due to time-dependent quantum spreading.  In general, it can be quite large, much larger than the classical entropy production, with a single $\mathcal{SE}$ zero-order state being the extreme case.  As we have seen, it is easy to  ``tune" wave packets between the classical limit of zero excess entropy production, and the limit of extreme entropy production by a single zero-order state.   One can speculate that $\Delta S^x$  represents excess expenditure of a kind of available or free energy.  In this connection, we have shown that excess entropy can be associated with distinctly nonclassical effects in quantum systems, such as heat flow from cold to hot and asymmetric temperature equilibration \cite{twobath}.  It could be very interesting to try to relate such effects to the quantitative treatment of excess entropy developed here.   

\appendix

\noindent {\bf Acknowledgment}  

This work benefitted from access to the Talapas supercomputer at the University of Oregon.

\bibliography{Suniv}
\bibliographystyle{unsrt}

\section*{Supplemental Information}

Here we present Supplemental Information to the main paper: ``On Quantum Entropy and Excess Entropy Production in a System-Environment Pure State."

\appendix
\section*{A: Analysis of the system-environment decomposition of the classical microcanonical entropy} \label{Q/T section} 
\stepcounter{section}

We show how a system-environment decomposition of the classical microcanonical Boltzmann entropy $S = k_B \ln W$ gives the standard result for the environment $\Delta S^\mathrm{micro}_\mathcal{E} = Q/T$ in Eq.~17 of the main text. 

The classical microcanonical ensemble is based on the idea of $W = \rho\delta E$ states in the microcanonical energy shell of width $\delta E$ with density of states $\rho$.  The entropy is given by Boltzmann's relation

\begin{equation} \label{suniv micro} S_{univ}^\mathrm{micro} = -\sum_{s,\epsilon} p_{s,\epsilon}^\mathrm{micro} \ln p_{s,\epsilon}^\mathrm{micro} = \ln W, \end{equation}

\noindent where 

\begin{equation} \label{p univ micro} p_{s,\epsilon}^\mathrm{micro} = \frac{1}{W}. \end{equation}  

\noindent The entropy can be decomposed into system and environment parts following Eq.~12 of the main text, 

\begin{equation} \label{suniv micro decomposition} S_{univ}^\mathrm{micro} = S_\mathcal{S}^\mathrm{micro} + S_\mathcal{E}^\mathrm{micro}.\end{equation}     

First consider the system component, analogous to $S_i$ in Eq.~12 of the main text,

\begin{equation}  \label{s sys micro def} S_\mathcal{S}^\mathrm{micro}  = -\sum p_s^\mathrm{micro} \ln p_s^\mathrm{micro}, \end{equation}

\noindent with system probabilities computed from Eq.~\ref{p univ micro} as

\begin{equation} \label{p sys micro} p_s^\mathrm{micro} = \sum_{\epsilon = 1}^{W_s} p_{s,\epsilon}^\mathrm{micro} = \frac{W_s}{W}, \end{equation}

\noindent where $W_s$ is the number of bath states $\epsilon$ that pair with $s$ in the microcanonical ensemble.  Using Eq.~\ref{p sys micro} to rewrite $\ln p_s^\mathrm{micro}$ in Eq.~\ref{s sys micro def} gives

\begin{equation} \label{s sys micro} S_\mathcal{S}^\mathrm{micro} = - \sum_s p_s^\mathrm{micro} (\ln W_s - \ln W) = \ln W- \sum_s p_s^\mathrm{micro} \ln W_s \end{equation}

Now consider the environment term

\begin{equation} \label{s env micro def} S_\mathcal{E}^\mathrm{micro} = \sum_s p_s^\mathrm{micro} \left(-\sum_\epsilon p_{\epsilon | s}^\mathrm{micro} \ln p_{\epsilon | s}^\mathrm{micro}  \right) \end{equation}

\noindent analogous to $\langle S_{\lambda }\rangle_{\{i\}}$ in Eq.~12 of the main text.  The conditional environment probabilities are calculated from Eq.~11 of the main text along with Eqs.~\ref{p univ micro} and \ref{p sys micro},

\begin{equation} \label{p epsilon micro} p_{\epsilon | s}^\mathrm{micro} = \frac{p_{s,\epsilon}^\mathrm{micro}}{p_s^\mathrm{micro}} = \frac{1}{W_s} \end{equation} 

\noindent Using Eq.~\ref{p epsilon micro} we can simplify the rightmost sum in Eq.~\ref{s env micro def}

\begin{equation} -\sum_\epsilon p_{\epsilon | s}^\mathrm{micro} \ln p_{\epsilon | s}^\mathrm{micro} = \ln W_s, \end{equation}

\noindent then putting this into Eq.~\ref{s env micro def} gives

\begin{equation} \label{s env micro} \langle S_\mathcal{E}^\mathrm{micro}\rangle_{\{sys\}} = \sum_s p_s^\mathrm{micro} \ln W_s . \end{equation}

\noindent The system and environment entropies Eqs.~\ref{s sys micro} and \ref{s env micro} clearly sum to the total microcanonical entropy $S_{univ}^\mathrm{micro}=\ln W$ in Eq.~\ref{suniv micro}, as needed.

We now consider a thermalization process where we begin with a constrained microcanonical ensemble of $W_0$ states for the initial state.  For example, this could correspond to a situation where the system begins in thermal isolation from the environment. The constraint is then removed, allowing heat to flow between the system and environment, resulting in a final state microcanonical ensemble with $W_f > W_0$.  The total entropy change is

\begin{equation} \label{delta suniv micro appendix} \Delta S_{univ}^\mathrm{micro} = \ln\frac{W_f}{W_0} = \ln\frac{\rho_f}{\rho_0} \end{equation}

\noindent The last equality comes from the microcanonical relation $W = \rho\delta E$ with $\rho$ the density of states in the microcanonical energy shell of width $\delta E$. The system entropy change from Eq.~\ref{s sys micro} is

\begin{equation} \label{delta s sys micro def} \Delta S_\mathcal{S}^\mathrm{micro} = \ln \frac{W_f}{W_0} - \sum_{s_f} p_{s_f}^\mathrm{micro} \ln W_{s_f} + \sum_{s_0} p_{s_0}^\mathrm{micro} \ln W_{s_0} \end{equation}

The system entropy change can be greatly simplified through a series of manipulations we will perform on the final two sums of Eq.~\ref{delta s sys micro def}.  This will lead to the final simple result for the system entropy in Eq.~\ref{delta s sys micro}, and will also be useful in deriving the environment entropy change in Eq.~\ref{delta s env micro}.  First, the sums can be combined by inserting the identities $\sum_{s_0} p_{s_0}^\mathrm{micro} = \sum_{s_f} p_{s_f}^\mathrm{micro} = 1$,

\begin{equation} \label{delta s sys micro intermediate} - \sum_{s_f} p_{s_f}^\mathrm{micro} \ln W_{s_f} + \sum_{s_0} p_{s_0}^\mathrm{micro} \ln W_{s_0} =  - \sum_{s_0,s_f} p_{s_0}^\mathrm{micro} p_{s_f}^\mathrm{micro} \ln \frac{W_{s_f}}{W_{s_0}}. \end{equation}

\noindent This is simplified by noting that for a heat bath environment $W \sim e^{E_\mathcal{E}/T}$, where $E_\mathcal{E} = E_{total}-E_s$ is the energy of the environment.  Then the ratio $W_{s_f}/W_{s_0}$ in the right of Eq.~\ref{delta s sys micro intermediate} can be expressed as

\begin{equation} \label{boltzmann factor} W_{s_f}/{W_{s_0}} = e^{-(E_{s_f} - E_{s_0})/T} \end{equation} 

\noindent where $E_{s_f} - E_{s_0} = \Delta E_\mathcal{S}$ is the energy difference between the final and initial system states $s_f$ and $s_0$.  Putting this into Eq.~\ref{delta s sys micro intermediate} gives

\begin{equation} -\sum_{s_0,s_f} p_{s_0}^\mathrm{micro} p_{s_f}^\mathrm{micro} \ln \frac{W_{s_f}}{W_{s_0}} =  \sum_{s_0,s_f} p_{s_0}^\mathrm{micro} p_{s_f}^\mathrm{micro}\frac{ E_{s_f} - E_{s_0} }{T} \end{equation}

\noindent Now we separate again into two terms

$$ \sum_{s_0,s_f} p_{s_0}^\mathrm{micro} p_{s_f}^\mathrm{micro}\frac{ E_{s_f} - E_{s_0} }{T} $$  
\begin{equation} = \frac{1}{T} \left( \sum_{s_0}  p_{s_0}^\mathrm{micro} \sum_{s_f} p_{s_f}^\mathrm{micro}E_{s_f} - \sum_{s_f}p_{s_f}^\mathrm{micro}\sum_{s_0} p_{s_0}^\mathrm{micro} E_{s_0} \right)  \end{equation}

\noindent Using the identities $\sum_s p_{s_0}^\mathrm{micro} = \sum_s p_{s_f}^\mathrm{micro} = 1$ this becomes

\begin{equation} \label{sys Delta E/T}  \frac{1}{T} \left( \sum_{s_f} p_{s_f}^\mathrm{micro}E_{s_f} - \sum_{s_0} p_{s_0}^\mathrm{micro} E_{s_0} \right) =  \frac{ \langle E_{\mathcal{S},f}\rangle - \langle E_{\mathcal{S},0}\rangle }{T} = \frac{\Delta \langle E_\mathcal{S}\rangle}{T} \end{equation}

\noindent Finally, we note that the system energy change is due solely to heat flow from the environment $\Delta \langle E_\mathcal{S}\rangle = -Q = -\Delta \langle E_\mathcal{E}\rangle$, so we can express the result in Eq.~\ref{sys Delta E/T} equivalently as

\begin{equation} \label{sys Q/T} \frac{\Delta \langle E_\mathcal{S}\rangle}{T} = \frac{-Q}{T}. \end{equation}

\noindent In total, Eqs.~\ref{delta s sys micro intermediate}-\ref{sys Q/T} show that

\begin{equation} \label{sum Q/T} - \sum_{s_f} p_{s_f}^\mathrm{micro} \ln W_{s_f} + \sum_{s_i} p_{s_0}^\mathrm{micro} \ln W_{s_0} = \frac{-Q}{T} \end{equation}

\noindent Putting this into Eq.~\ref{delta s sys micro def} the system entropy change takes the standard and simple final form

\begin{equation} \label{delta s sys micro} \Delta S_\mathcal{S}^\mathrm{micro} = \ln \frac{W_f}{W_0} - \frac{Q}{T}. \end{equation} 

Now consider the entropy change of the environment. From the basic relation of Eq.~\ref{s env micro} this is

\begin{equation} \Delta S_\mathcal{E}^\mathrm{micro}  = \sum_{s_f} p_{s_f}^\mathrm{micro} \ln W_{s_f} - \sum_{s_0} p_{s_0}^\mathrm{micro} \ln W_{s_0} . \end{equation}

\noindent Using Eq.~\ref{sum Q/T} this is simply

\begin{equation} \label{delta s env micro} \Delta S_\mathcal{E}^\mathrm{micro} = \frac{Q}{T}, \end{equation}

\noindent which is the standard thermodynamic result.  Note this is an exact equality for a standard heat bath with the level density behavior of Eq.~\ref{boltzmann factor}.  Thus we have demonstrated $\Delta S_\mathcal{E}^\mathrm{micro} = Q/T$, as stated in Eq.~17 of the main text.

\section*{B: Lorentzian state distributions} \label{lorentzian appendix}
\stepcounter{section}
In this appendix we show how we obtain the time-evolving Lorentzian states discussed in Section VI of the main text.  The time-evolution of an initial state $\vert \Psi(0)\rangle$ follows the Schr\"odinger equation, expressed in terms of the eigenstates $\vert \xi\rangle$ as

\begin{equation} \label{schrodinger equation} \vert \Psi(t) \rangle = e^{-i \hat H t} \vert \Psi(0)\rangle = \sum_{\xi} c_\xi e^{-i E_\xi t} \vert \xi \rangle. \end{equation}

\noindent Our approach will be to first analyze the structure of the eigenstates $\vert \xi \rangle$, then use the eigenstate structure to analyze the time-dependent behavior of the $\vert s\rangle\vert \epsilon\rangle$ and Lorentzian initial states.

Some of our important results for the average equilibrium behavior of time-dependent states, Eqs.~\ref{typical Lorentzian eigenstate} and \ref{se eq} below, were obtained in nearly the same form by Deutsch in his well-known paper of 1991 \cite{Deutsch1991,DeutschSupplemental} where he developed the ideas behind the eigenstate thermalization hypothesis approach to quantum thermodynamics \cite{Deutsch}.  Our model varies somewhat from Deutsch's, so that our eigenstates require an additional parameter  in Eq.~\ref{typical Lorentzian eigenstate} that was not included in Deutsch's work.  The widths of the eigenstates in from our calculations also vary from Deutsch's result by a factor of two, in agreement with a recent re-evaluation of Deutsch's work by Nation and Porras in Ref.~\cite{Porras}.

In addition to analyzing the average equilibrium behavior of time-dependent states, we also analyze the fluctuations of a time-evolving state about its average and to develop the idea of a time-evolving initial Lorentzian state.  This provides the critical relations in Eq.~27 of the main, appearing again in this appendix as Eqs.~\ref{psi se eq} and \ref{Psi eq Lorentzian}.

\subsection{Eigenstates}

We build up to an analysis of time-dependent states beginning with the structure of the eigenstates $\vert \xi \rangle$, for the system of a few energy levels, the environment with an exponential density of states, and a random-matrix system-environment interaction, as described in the main text and Ref.~\cite{micro}. In the system-environment zero-order energy basis $\{\vert s\rangle\vert \epsilon\rangle\}$ the eigenstates are expressed as

\begin{equation} \label{eigenstate} \vert \xi \rangle = \sum_{s,\epsilon} c_{s,\epsilon}^{(\xi)} \vert s \rangle\vert \epsilon\rangle, \end{equation}

\noindent where the coefficients $c_{s,\epsilon}^{(\xi)}$ are real numbers since the Hamiltonian is real.  

Deutsch \cite{Deutsch1991,DeutschSupplemental} and Nation and Porras \cite{Porras} derived the eigenstate coefficients $c_{s,\epsilon}^{(\xi)}$ in a very similar model with a random interaction between evenly spaced system-environment levels.  We find that our eigenstates can be very well fit by their result with the addition of a fit parameter $\Delta E_0$.  It seems likely to us that this is related to the exponential level density in our environment as opposed to the evenly spaced levels they considered.  With this additional fit parameter, our eigenstates coefficients can be described statistically as

\begin{equation} \label{typical Lorentzian eigenstate} c_{s,\epsilon}^{(\xi)} \approx g_{s,\epsilon}^{(\xi)} \sqrt{L_\xi(E_s + E_\epsilon)},\end{equation}

\noindent where $L_\xi(E_s + E_\epsilon)$ is a Lorentzian distribution and the $g_{s,\epsilon}^{(\xi)}$ give random variations about the Lorentzian average.   The Lorentzian is

\begin{equation} \label{Lorentzian} L_\xi(E_s + E_\epsilon) = \frac{1}{\pi}\frac{\gamma_\xi/\rho(E_\xi)}{(E_\xi -E_s - E_\epsilon - \Delta E_0)^2 + \gamma_\xi^2 },\end{equation} 

\noindent with half-width at half-max

\begin{equation} \label{gamma xi} \gamma_\xi = \pi k^2 \rho(E_\xi), \end{equation}

\noindent where $E_\xi$ is the eigenstate energy, $\rho(E_\xi)$ is the total density of system-environment zero-order states, and $\Delta E_0$ is a fit parameter that sets the center of the Lorentzian.  The small parameter $\Delta E_0$ varies slightly between eigenstates, but we will approximate it as constant here to simplify the analysis, finding that this is entirely adequate for describing our results. 

   \begin{figure}[h]
\begin{center}
\includegraphics[width=8cm,height=11cm,keepaspectratio,angle=0]{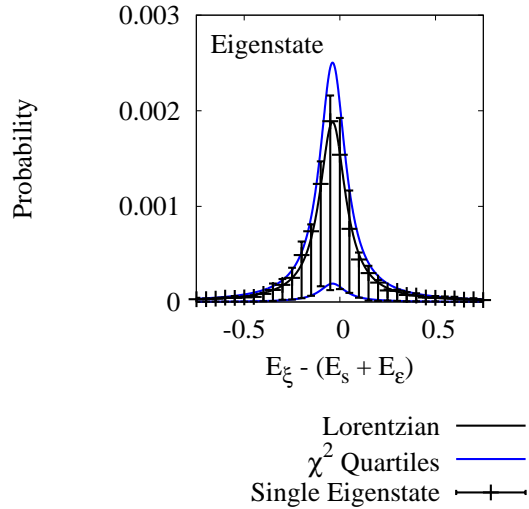}
\end{center}
\caption[Single eigenstate Lorentzian distribution.]{Average squared coefficients $\langle |c_{s,\epsilon}^{(\xi)}|^2 \rangle$ for a single eigenstate follow the Lorentzian distribution of Eq.~\ref{Lorentzian}.  Error bars show the first and third quartiles of the distribution of $|c_{s,\epsilon}^{(\xi)}|^2$ in each data point.  The quartiles of the coefficient distributions are in good agreement with the quartiles of the single degree of freedom $\chi^2$ distribution with the Lorentzian mean, shown in blue.}
\label{eigenstate fig}
\end{figure}

Fig.~\ref{eigenstate fig} shows an example of a single eigenstate calculated by exact diagonalization of our Hamiltonian.  In the figure, we averaged squared-coefficients $|c_{s,\epsilon}^{(\xi)}|^2$ with nearby energies $E_s + E_\epsilon$ to get the averages $\langle |c_{s,\epsilon}^{(\xi)}|^2 \rangle$ shown as data points in the figure.  The averages $\langle |c_{s,\epsilon}^{(\xi)}|^2 \rangle$ represent the average probability of measuring an $\vert s \rangle\vert \epsilon\rangle$ state of energy $E_s + E_\epsilon$ when in the eigenstate $\vert \xi\rangle$ of energy $E_{\xi}$.  The averages are very well described by the Lorentzian $\langle |c_{s,\epsilon}^{(\xi)}|^2 \rangle \approx L_\xi(E_s + E_\epsilon)$, with $L_\xi$ from Eq.~\ref{Lorentzian}.

\begin{figure}[h]
\begin{center}
\includegraphics[width=8cm,height=11cm,keepaspectratio,angle=0]{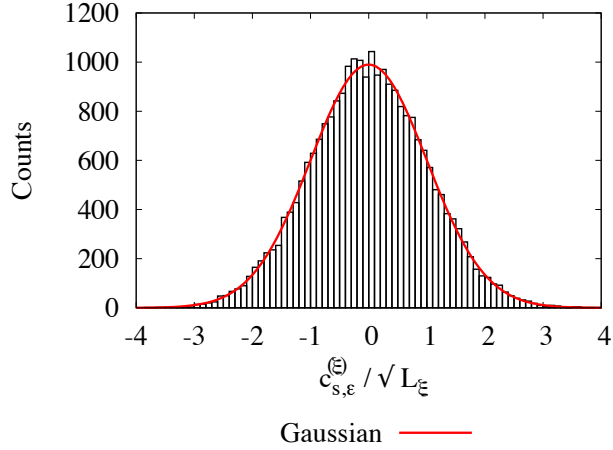}
\end{center}
\caption[Coefficient variations for an eigenstate.]{Histogram counts of coefficient variations for an eigenstate.  The variations $g_{s,\epsilon}^{(\xi)} = c_{s,\epsilon}^{(\xi)}/\sqrt{L_\xi}$ in the single eigenstate coefficients of Eq.~\ref{typical Lorentzian eigenstate} follow the standard Gaussian distribution of Eq.~\ref{Gaussian variate}.}
\label{eigenstate coeff fig}
\end{figure}

 The asymmetric error bars in the figure show the first and third quartiles of the distribution of squared coefficients $|c_{s,\epsilon}^{(\xi)}|^2$ for each data-point average $\langle |c_{s,\epsilon}^{(\xi)}|^2 \rangle$. The quartiles are in good agreement with the quartiles of a single degree of freedom $\chi^2$ distribution with mean $L_\xi(E_s+E_\epsilon)$.  The $\chi^2$ distribution describes a sum of squared random Gaussian variates.  This suggests that the $g_{s,\epsilon}^{(\xi)}$ in Eq.~\ref{typical Lorentzian eigenstate} behave as random standard Gaussian variates, so that the squared coefficients $|c_{s,\epsilon}^{(\xi)}|^2$ follow the $\chi^2$ distribution with mean $L_\xi$.  To check this, in Fig.~\ref{eigenstate coeff fig} we plot the distribution of the $g_{s,\epsilon}^{(\xi)} = c_{s,\epsilon}^{(\xi)}/\sqrt{L_\xi(E_s + E_\epsilon)}$, where they are indeed seen to follow a standard Gaussian distribution

\begin{equation} \label{Gaussian variate} p(g_{s,\epsilon}^{(\xi)}) = p\left(\frac{c_{s,\epsilon}^{(\xi)}}{\sqrt{L_\xi(E_s + E_\epsilon)}}\right) \sim e^{-{g_{s,\epsilon}^{(\xi)}}^2/2}. \end{equation} 

\noindent The Gaussian variations for $g_{s,\epsilon}^{(\xi)}$ in Eq.~\ref{Gaussian variate} explain the $\chi^2$ distributed quartiles in Fig.~\ref{eigenstate fig}, and are consistent the work of Deutsch \cite{Deutsch1991,DeutschSupplemental} and Nation and Porras \cite{Porras}.   We have thus arrived at the description of the eigenstates in Eq.~\ref{typical Lorentzian eigenstate}, with the Lorentzian $L_\xi$ of Eq.~\ref{Lorentzian} and the random variations $g_{s,\epsilon}^{(\xi)}$ of Eq.~\ref{Gaussian variate}.

The Gaussian fluctuations in the basis state probabilities of Eq.~\ref{typical Lorentzian eigenstate} are related to the random structure of the interaction, as discussed by Deutsch \cite{Deutsch1991,DeutschSupplemental}.  They also have a connection to the random states considered in the ``typicality" approaches to quantum statistical mechanics \cite{Goldstein2006,Goldstein2010,vNcommentary,Reimann2008,Popescu2009,Popescu2006,Reimann2016,Goldstein2015}.  These approaches seek to rationalize thermalization behavior by analyzing the statistics of random states.  An unbiased sampling of random states is accomplished by taking coefficients as random Gaussian variates \cite{Goldstein2006}, similar to our Eq.~\ref{Gaussian variate}.  Here, the eigenstates can be thought of as random or ``typical" states within their Lorentzian windows, as seen in Fig.~\ref{eigenstate coeff fig}.

\subsection{Time evolution of an $\vert s \rangle \vert \epsilon\rangle$ initial state} \label{se state section}\vspace{24pt}

Our goal in this section is to understand the behavior of a very simple time-dependent state, from Eq.~26 of the main text, that begins in single zero-order basis state 

\begin{equation} \label{Psi0 se} \vert \Psi_{s,\epsilon}(t) \rangle = e^{-i\hat H t}\vert s \rangle\vert \epsilon\rangle = \sum_{s',\epsilon'} c_{s',\epsilon'}(t) \vert s'\rangle \vert \epsilon'\rangle. \end{equation}  

\noindent  Our analysis will give the Lorentzian behavior for the time-evolved  state $\vert \Psi_{s,\epsilon}(t)\rangle$ seen in Fig.~3 and Eqs.~27-29 of the main text.  We now briefly describe these results, repeated here in Eqs.~\ref{psi se eq}-\ref{gamma se}, before going into the mathematical details of how we obtain the results in the remainder of the section.

   \begin{figure}[h]
\begin{center}
\includegraphics[width=8cm,height=11cm,keepaspectratio,angle=0]{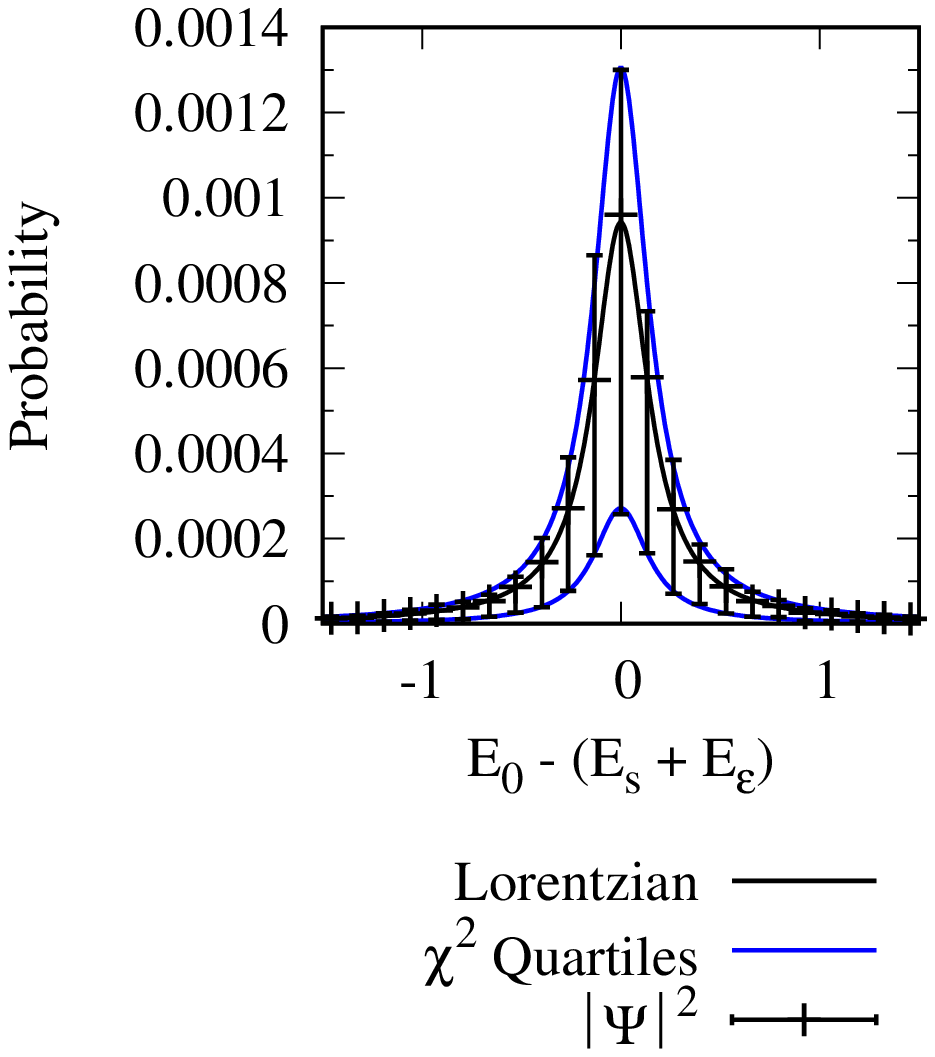}
\end{center}
\caption[Time-evolved \se initial state follows a Lorentzian.]{Average squared coefficients $\langle |c_{s',\epsilon'}(t)|^2 \rangle$ for a time-evolved $\vert s \rangle\vert \epsilon\rangle$ state of Eq.~\ref{Psi0 se} follow a Lorentzian distribution with twice the width of the eigenstates.  Error bars show the first and third quartiles of the distribution of the individual $|c_{s',\epsilon'}(t)|^2$ in each data point.  The quartiles of the coefficient distributions are in good agreement with the quartiles of the two degree of freedom $\chi^2$ distribution scaled by half the Lorentzian weight, shown in blue.}
\label{se state fig}
\end{figure}

Fig.~\ref{se state fig} shows an example of average squared coefficients $\langle|c_{s',\epsilon'}(t)|^2\rangle$ for a time-evolved $\vert s \rangle\vert \epsilon\rangle$ state of the type in Eq.~\ref{Psi0 se}, at a time $t$ at equilibrium (the results are similar for other choices of $t$).  The state is the same as in Fig.~3 of the main text.  
The average squared coefficients $\langle |c_{s,\epsilon}|^2\rangle$ for the state follow a Lorentzian distribution with twice the width of the eigenstate seen previously in Fig.~\ref{eigenstate fig}; note the energy range in Fig.~\ref{se state fig} is doubled relative to Fig.~\ref{eigenstate fig}.  The equilibrated state of Eq.~\ref{Psi0 se} can be expressed in the Lorentzian form from Eqs.~27-29 of the main text,

\begin{equation} \label{psi se eq} \vert \Psi_{s,\epsilon}(t)\rangle \approx \sum_{s',\epsilon'} \tilde g_{s',\epsilon'}  \sqrt{L^{(s,\epsilon)}_f} \vert s'\rangle\vert\epsilon'\rangle , \end{equation}

\noindent where the $\tilde g_{s',\epsilon'}$ are random complex fluctuations and $L^{(s,\epsilon)}_f$ is a Lorentzian centered at the initial $\vert s \rangle \vert \epsilon\rangle$ energy $E_0 = E_s + E_\epsilon$,

\begin{equation} \label{Lorentzian se} L^{(s,\epsilon)}_f(E_{s'} + E_{\epsilon'} ) = \frac{1}{\pi}\frac{\gamma_{f} /\rho_f}{(E_{s'} + E_{\epsilon'} - E_0)^2 + \gamma_{f} ^2 },\end{equation} 

\noindent with half-width at half-max

\begin{equation} \label{gamma se} \gamma_{f} = 2\pi k^2 \rho_f, \end{equation}

\noindent where $\rho_f = \rho(E_0)$ is the total density of system-environment states at $E_0$.  The Lorentzian is similar to the eigenstate Lorentzian in Eqs.~\ref{Lorentzian} and \ref{gamma xi}, except that the eigenstate energy has been replaced with the initial state energy $E_0 = E_s + E_\epsilon$, the width is doubled to $2\pi k^2 \rho_f$, and there is no median energy parameter $\Delta E_0$. 

The error bars in Fig.~\ref{se state fig} show the quartiles of the distributions of squared coefficients for each data point, they are in very good agreement with the quartiles of a two degree of freedom $\chi^2$ distribution scaled by 1/2 the Lorentzian.  This will be related to the structure of the random deviation terms $\tilde g_{s',\epsilon'}$ in Eq.~\ref{psi se eq}, which we will find to follow statistics where their real and imaginary parts can be treated as random Gaussian variates, as in Eq.~24 of the main text.  We now discuss how we obtain these results mathematically.

\subsubsection{Average Lorentzian Distribution $\bm{L_{s,\epsilon}}$ for the time-evolved $\bm{\vert s \rangle \vert \epsilon\rangle}$ state} \label{se avg section}\vspace{24pt}

To derive the average Lorentzian behavior of Eq.~\ref{psi se eq}, we begin by calculating the average equilibrium state distribution for the time-evolving state of Eq.~\ref{Psi0 se}.  The average equilibrium behavior is given by the long-time average of the density operator

\begin{equation} \label{time avg} \langle\hat \rho_{s,\epsilon}\rangle_{t\to\infty} = \big\langle\vert \Psi_{s,\epsilon}(t) \rangle\langle \Psi_{s,\epsilon}(t)\vert\big\rangle_{t\to\infty} = \sum_{\xi,\xi'} c^{(s,\epsilon)}_\xi {c^{(s,\epsilon)}_{\xi'}}^* \big\langle e^{-i(E_\xi-E_{\xi'})t}\big\rangle_{t\to\infty} \vert \xi \rangle \langle \xi'\vert, \end{equation}

\noindent where the time-averages are $\langle x \rangle_{t\to\infty} \equiv \lim_{t\to\infty} (1/t) \int_0^t d\tau x(\tau)$ and the coefficients are given by Eq.~\ref{typical Lorentzian eigenstate} with $c_\xi^{(s,\epsilon)} = \langle \xi \vert s \rangle \vert \epsilon \rangle = \ {c_{s,\epsilon}^{(\xi)}}^*$.  The energy eigenvalues are non-degenerate since there are no symmetries in the random matrix model, so the cross terms average to zero and 

\begin{equation} \label{se time avg} \langle\hat \rho_{s,\epsilon}\rangle_{t\to\infty} =  \sum_{\xi} |c^{(s,\epsilon)}_\xi|^2 \vert \xi \rangle \langle \xi\vert  = \sum_\xi |g_{s,\epsilon}^{(\xi)}|^2 L_\xi(E_0) \vert \xi \rangle \langle \xi\vert ,\end{equation}

\noindent where the last equality has replaced the $ |c^{(s,\epsilon)}_\xi|^2$ with the expressions from Eq.~\ref{typical Lorentzian eigenstate}, with the initial state energy $E_0 = E_s + E_\epsilon$.  

We are interested in the distribution of the time-average density operator of Eq.~\ref{se time avg} in the $\{\vert s \rangle\vert \epsilon\rangle\}$ basis, where the diagonal elements are  $\langle s' \vert \langle \epsilon'\vert  \langle\hat \rho_{s,\epsilon}\rangle_{t\to\infty}  \vert \epsilon'\rangle \vert s'\rangle$.  Using the form of the eigenstates in Eqs.~\ref{eigenstate} and \ref{typical Lorentzian eigenstate} the diagonal elements of the density operator in the zero-order basis are

\begin{equation} \langle s' \vert \langle \epsilon'\vert  \langle\hat \rho_{s,\epsilon}\rangle_{t\to\infty}  \vert \epsilon'\rangle \vert s'\rangle \approx  \sum_{\xi} |g_{s,\epsilon}^{(\xi)}|^2 |g_{s',\epsilon'}^{(\xi)}|^2 L_\xi(E_0)  L_\xi(E_{s'} + E_{\epsilon'}). \end{equation}

\noindent We assume that the Gaussian variates are statistically independent from the Lorentzian factors so we can simply approximate them with their mean values $\langle |g_{s,\epsilon}^{(\xi)}|^2 \rangle = \langle |g_{s',\epsilon'}^{(\xi)}|^2 \rangle = 1$ for all values of the indices $\xi,s,\epsilon,s',\epsilon'$.  With this approximation

\begin{equation}\label{rho se mid} \langle s' \vert \langle \epsilon'\vert  \langle\hat \rho_{s,\epsilon}\rangle_{t\to\infty}  \vert \epsilon'\rangle \vert s'\rangle \approx  \sum_{\xi}  L_\xi(E_0)  L_\xi(E_{s'} + E_{\epsilon'}). \end{equation}

We will now make two approximations to greatly simplify this sum, resulting ultimately in a single Lorentzian factor.  The first approximation uses a single density of states $\rho(E_\xi) = \rho(E_0)$ evaluated at the initial state energy $E_0 = E_s + E_\epsilon$ instead of the variable eigenstate energy $E_\xi$.  This approximation is reasonable since most of the sum comes from eigenstates with eigenenergies $E_\xi \approx E_0$ where the Lorentzians are near their maxima in Eq.~\ref{Lorentzian}.  The second approximation is to replace the sum by an integral over all energies, which is reasonable since the discrete energy level spacings are small.  With these approximations Eq.~\ref{rho se mid} becomes

$$ \langle s' \vert \langle \epsilon'\vert  \langle\hat \rho_{s,\epsilon}\rangle_{t\to\infty}  \vert \epsilon'\rangle \vert s'\rangle$$
\begin{equation} \label{convolution integral} \approx  \frac{1}{\pi^2\rho} \int_{-\infty}^{\infty} dE_\xi \frac{\pi k^2 \rho}{(E_0 + \Delta E_0 - E_\xi)^2 +(\pi k^2 \rho)^2} \frac{\pi k^2 \rho}{(E_\xi - E_{s'} - E_{\epsilon'} - \Delta E_0 )^2 + (\pi k^2 \rho)^2},  \end{equation} 

\noindent where $\rho = \rho(E_0)$.  There is an additional factor of the density of states $\rho(E_0)$ in the integrand of Eq.~\ref{convolution integral} in comparison to the summands in Eq.~\ref{rho se mid} since there are $\rho dE_\xi$ summands within each interval $dE_\xi$ of the integration. The integral gives the convolution of two Lorentzians. We evaluated the integral using {\it Mathematica}, the result is a Lorentzian with twice the half-width at half-max and a central energy at $E_{s'} + E_{\epsilon'} = E_0$,

 \begin{equation} \label{se eq}  \langle s' \vert \langle \epsilon'\vert  \langle\hat \rho_{s,\epsilon}\rangle_{t\to\infty}  \vert \epsilon'\rangle \vert s'\rangle \approx  \frac{1}{\pi} \frac{2\pi k^2}{(E_{s} + E_{\epsilon'} - E_0)^2 + (2\pi k^2 \rho)^2}. \end{equation}

The relation Eq.~\ref{se eq} gives the average Lorentzian in Eq.~\ref{Lorentzian se}  and Fig.~\ref{se state fig} at the start of this section and in Eqs.~27-29 and Fig.~3 of the main text.  It is a Lorentzian centered at the initial state energy $E_0 = E_s + E_\epsilon$, with twice the width of the eigenstates, obtained through the convolution of Lorentzians in Eq.~\ref{convolution integral}.  This result was also obtained by Deutsch \cite{Deutsch1991,DeutschSupplemental} in a similar model.  We will now consider the fluctuations about this Lorentzian average, to determine the factors $\tilde g_{s',\epsilon'}$ in Eq.~\ref{psi se eq}. 

\subsubsection{Fluctuations $\bm{\tilde g_{s',\epsilon'}}$ in the coefficients of the time-evolved $\bm{\vert s \rangle \vert \epsilon\rangle}$ state}\label{se state fluctuation section}

Now we would like to consider the time-dependent state Eq.~\ref{Psi0 se} as undergoing fluctuations about its equilibrium Lorentzian average from Eq.~\ref{se eq}, where the fluctuations are given by the factors $\tilde g_{s',\epsilon'}$ in Eq.~\ref{psi se eq}.  We expect that the average squared fluctuation is unity, $\langle |\tilde g_{s',\epsilon'}|^2 \rangle = 1$, so that the $|c_{s',\epsilon'}(t)|^2$ follow the Lorentzian on average. We also expect that the real and imaginary parts of $\tilde g_{s',\epsilon'}$ should contribute equally on average.  This implies that the fluctuation term can be expressed as

\begin{equation} \label{g decomposition} \tilde g_{s',\epsilon'} = \frac{g_{s',\epsilon'} + ig'_{s',\epsilon'}}{\sqrt{2}}, \end{equation}

\noindent where the real and imaginary components $g_{s',\epsilon'}$ and $g'_{s',\epsilon'}$ each have the average squared values $\langle g_{s,\epsilon}^2\rangle  = \langle{g'_{s,\epsilon}}^2\rangle = 1$, so that $\langle |\tilde g_{s',\epsilon'}|^2\rangle=1$.

 We examine the real and imaginary  components $g_{s',\epsilon'}$ and $g'_{s',\epsilon'}$ separately, in comparison with the exact coefficients  $c_{s',\epsilon'}(t)$ of Eq.~\ref{Psi0 se}.  By comparison of Eqs.~\ref{Psi0 se}, \ref{psi se eq}, and \ref{g decomposition}, we have the following relations for $g_{s',\epsilon'}$ and $g'_{s',\epsilon'}$,

\begin{equation} \label{re g} g_{s',\epsilon'} = \mathrm{Re}\left( \frac{ \tilde g_{s',\epsilon'}}{1/\sqrt{2}}\right) = \mathrm{Re} \left(\frac{c_{s',\epsilon'}(t)}{\sqrt{L^{(s,\epsilon)}_f(E_{s'} + E_{\epsilon'})/2}}\right) \end{equation}

\noindent and

\begin{equation} \label{im g} g_{s',\epsilon'}' =  \mathrm{Im}\left( \frac{ \tilde g_{s',\epsilon'}}{1/\sqrt{2}}\right)  = \mathrm{Im} \left(\frac{c_{s',\epsilon'}(t)}{\sqrt{L^{(s,\epsilon)}_f(E_{s'} + E_{\epsilon'})/2}}\right). \end{equation}

\begin{figure*}
\begin{center}
\includegraphics[width=8cm,height=11cm,keepaspectratio,angle=0]{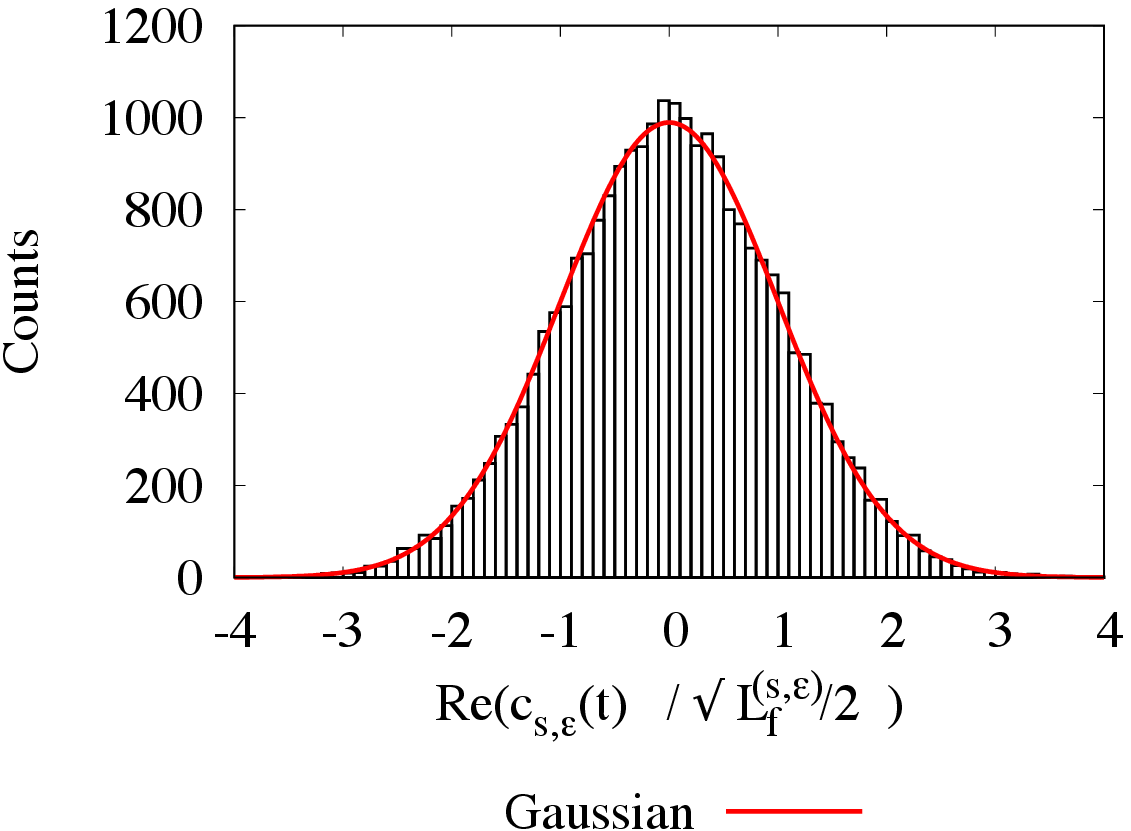}
\includegraphics[width=8cm,height=11cm,keepaspectratio,angle=0]{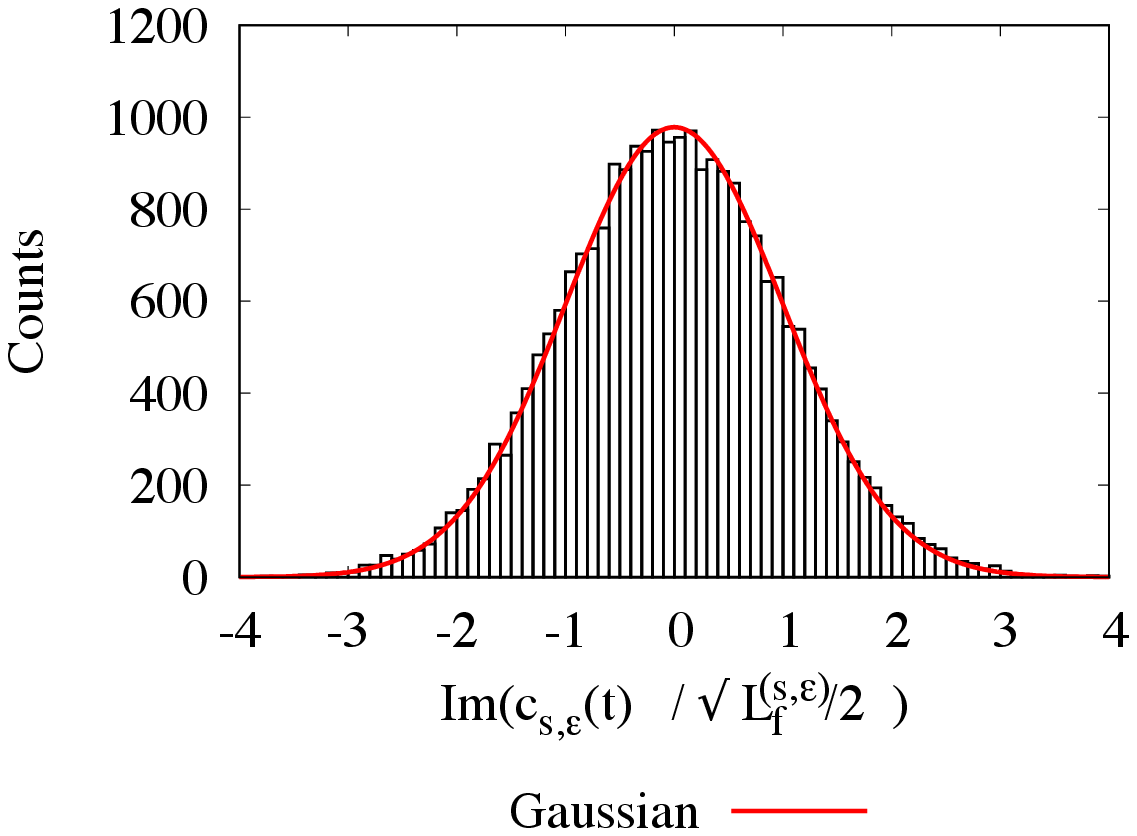}
\end{center}
\caption[Coefficient variations for a time-evolved \se initial state.]{Histogram counts of coefficient variations for a time-evolved \se initial state.  The real and imaginary parts the variations in Eqs.~\ref{re g} and \ref{im g} each follow a Gaussian distribution. }
\label{se state coeff fig}
\end{figure*}

Fig.~\ref{se state coeff fig} shows the distributions of the $g_{s',\epsilon'}$ and $g_{s',\epsilon'}' $ taken as the right hand sides of Eqs.~\ref{re g} and \ref{im g}, at an instant in time $t$ after the $\vert s \rangle \vert \epsilon\rangle$ initial state has evolved to equilibrium (the results are similar for other $t$).  The $g_{s,\epsilon}$ and $g_{s,\epsilon}' $ each follow standard Gaussian distributions, indicating that they are each distributed as random Gaussian variates as in Eq.~\ref{Gaussian variate}.  This is also consistent with the quartile distributions observed previously in Fig.~\ref{se state fig}; the squared complex fluctuation term $|\tilde g_{s',\epsilon'}|^2 = (g_{s,\epsilon}^2 + {g_{s,\epsilon}'}^2)/2$ has the distribution of a sum of two squared random standard Gaussian variates, which gives coefficients that follows the two degree of freedom $\chi^2$ distribution scaled by $L^{(s,\epsilon)}_f$/2, with the quartiles shown in Fig.~\ref{se state fig}.  Thus, taking $\tilde g_{s',\epsilon'}$ as the sum Eq.~\ref{g decomposition} with $g_{s,\epsilon} $ and $g'_{s,\epsilon}$ as random Gaussian variates is giving an entirely consistent description of our results in Figs.~\ref{se state fig} and \ref{se state coeff fig}.  

This completes our analysis of the time-evolved $\vert s \rangle\vert \epsilon\rangle$ state in Eq.~\ref{psi se eq}, where the coefficients $c_{s',\epsilon'}(t)$ are given in terms of the the Lorentzian averages $\sqrt{L_f^{(s,\epsilon)}}$ determined by the analysis of the last section and the complex Gaussian variate fluctuation terms $\tilde g_{s',\epsilon'}$ we have just discussed.

\subsection{Time evolution of a Lorentzian initial state} \label{Lorentzian state section appendix}\vspace{24pt}

Now we consider the time evolution of Lorentzian initial states, as in Fig.~2, Eqs.~22-24, and Eqs.~27-29 of the main text.  Our goal here is to systematically characterize the equilibration behavior of the Lorentzian initial states, similar to what we did in Section \ref{se state section} for the $\vert s \rangle \vert \epsilon \rangle$ initial states. 

We consider the initial Lorentzian state from  Eqs.~22-23 of the main text, repeated here as

\begin{equation}\label{Psi0 Lorentzian} \vert \Psi_{L}^0 \rangle = \sum_{\epsilon} \tilde g_{s,\epsilon} \sqrt{L_0} \vert s \rangle \vert \epsilon\rangle, \end{equation}

\noindent where the $\tilde g_{s,\epsilon}$ are random complex Gaussian variates as in Eq.~\ref{g decomposition} and $L_0$ is the initial state Lorentzian

\begin{equation} \label{L0} L_0(E_s + E_\epsilon) = \frac{1}{\pi} \frac{\gamma_0/\rho_0}{(E_s + E_\epsilon - E_0)^2 + \gamma_0^2}, \end{equation}

\noindent where $\gamma_0$ is the half-width at half-max, $E_0$ is the central Lorentzian energy, and $\rho_0(E_0)$ is the density of system-environment states with the system in its initial state $\vert s \rangle$ at the initial state energy $E_0 = E_s + E_\epsilon$.  Our goal is to show that this evolves into the final equilibrium state Lorentzian from Eqs.~27-29 of the main text, repeated here as

\begin{equation} \label{Psi eq Lorentzian} \vert \Psi_{L}^f (t) \rangle = \sum_{s,\epsilon} \tilde g_{s,\epsilon} \sqrt{L_f} \vert s \rangle \vert \epsilon\rangle, \end{equation}

\noindent with the final state Lorentzian

\begin{equation} \label{L0 final} L_f (E_s+E_\epsilon) = \frac{1}{\pi} \frac{\gamma_f/\rho_f}{(E_s + E_\epsilon - E_0)^2 + \gamma_f^2}, \end{equation}

\noindent with half-width at half-max

\begin{equation} \gamma_f = \gamma_0 + 2\pi k^2 \rho_f, \end{equation}

\noindent where $\rho_f = \rho(E_0)$ is the total density of system-environment zero-order states (when all system levels are accessible at equilibrium).

\begin{figure*}
\begin{center}
\includegraphics[width=7cm,height=11cm,keepaspectratio,angle=0,trim={2cm 0 0 0},clip]{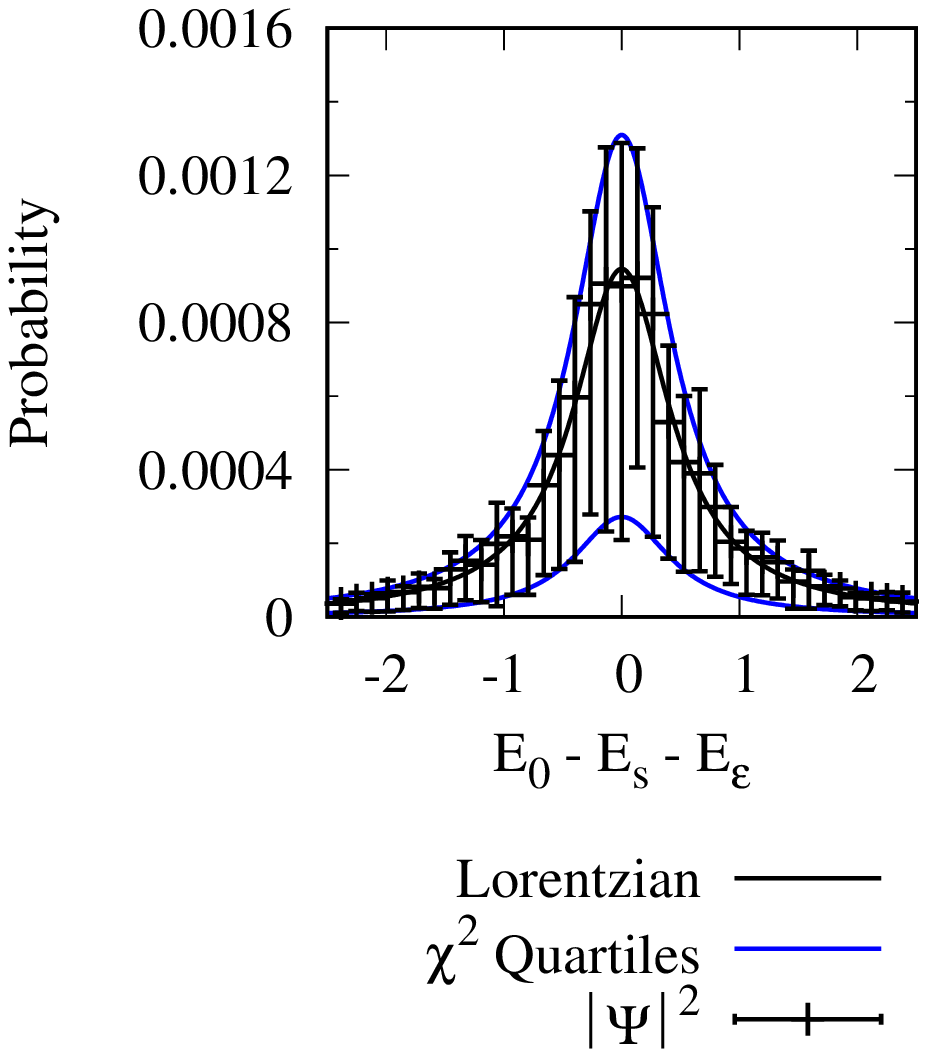}
\includegraphics[width=7cm,height=11cm,keepaspectratio,angle=0,trim={2cm 0 0 0},clip]{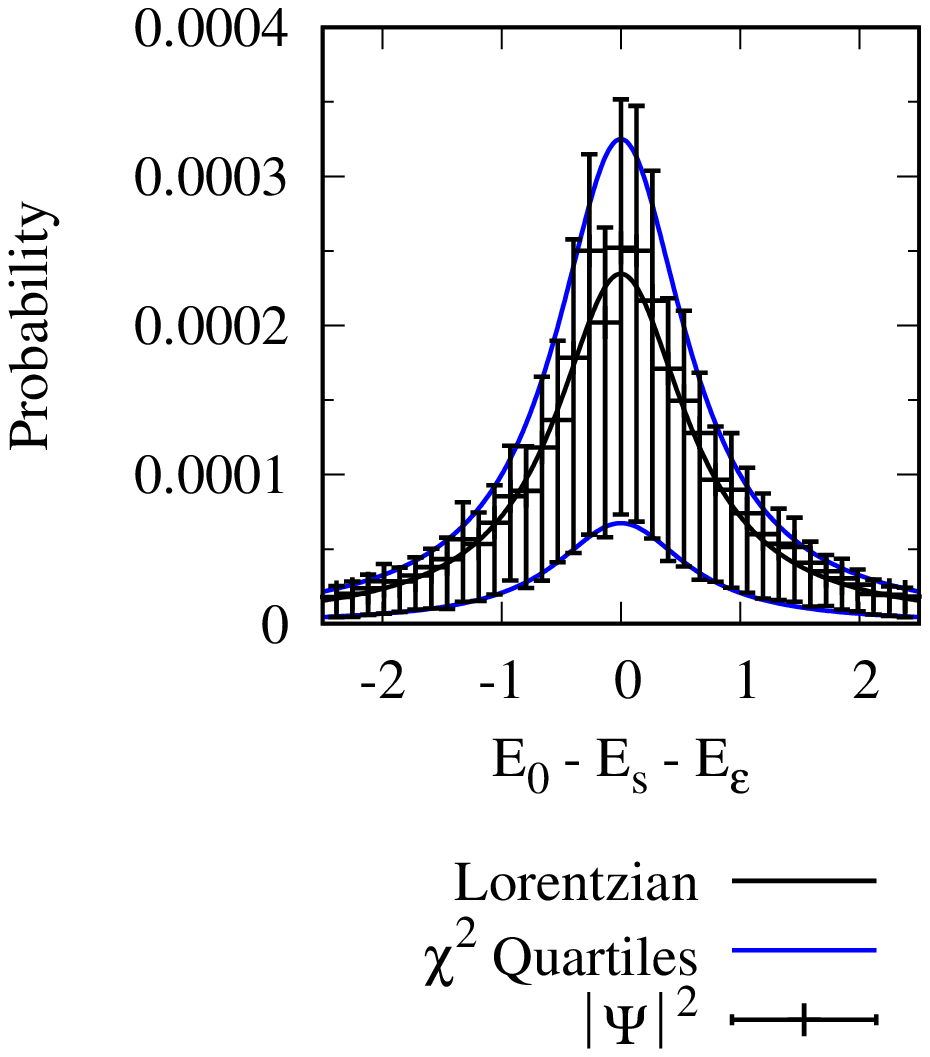}
\end{center}
\caption[Time-evolved Lorentzian initial state follows a Lorentzian.]{Average squared coefficients $\langle |c_{s',\epsilon'}(t)|^2 \rangle$ for a time-evolved Lorentzian initial state of Eq.~\ref{Psi0 Lorentzian} follows a Lorentzian distribution with increased width (the state is the same as in Fig.~\ref{superdynamics}).  Error bars show the first and third quartiles of the distribution of the individual $|c_{s',\epsilon'}(t)|^2$ in each data point.  The quartiles of the coefficient distributions are in good agreement with the quartiles of the two degree of freedom $\chi^2$ distribution scaled by half the Lorentzian weight, shown in blue.}
\label{Lorentzian state fig}
\end{figure*}

Fig.~\ref{Lorentzian state fig} shows an initial Lorentzian state of Eq.~\ref{Psi0 Lorentzian} on the left and a time-evolved version of the same state as in Eq.~\ref{Psi eq Lorentzian} on the right.  The state is the same as in Fig.~2 of the main text.  The final state Lorentzian $L_f$ is similar to the initial state Lorentzian $L_0$ except the width $\gamma_f$ is increased by twice the approximate widths of the eigenstates $2 \pi k^2 \rho_f$.  To rationalize this behavior, we will begin by analyzing the average equilibrium behavior of the time-evolving initial Lorentzian state, then analyze the fluctuations about the average to determine the $\tilde g_{s,\epsilon}$.  The fluctuations will follow the same type of random Gaussian structure as we had for the time-evolved \se states, giving the blue $\chi^2$ quartiles in the figure.

\subsubsection{Average final Lorentzian distribution for the time-evolved Lorentzian initial state}

To determine the average final state Lorentzian in Eq.~\ref{Psi eq Lorentzian}, we will calculate the average equilibrium behavior of the time-evolving Lorentzian initial state of Eq.~\ref{Psi0 Lorentzian}, analogous to what we did with the average time-evolving $\vert s\rangle \vert \epsilon\rangle$ initial state in Section \ref{se avg section}.  To begin, express the initial state density operator as

\begin{equation} \label{Lorentzian density op initial} \hat \rho_{L}^0 = \hat \rho_{L}^{0,\mathrm{diag}} + \hat \rho_{L}^{0,\mathrm{coh}} ,\end{equation}

\noindent where $\hat \rho_{L}^{0,\mathrm{diag}}$ gives the diagonal component with trace of unity,

\begin{equation} \hat \rho_{L}^{0,\mathrm{diag}} = \sum_{s,\epsilon} |\tilde g_{s,\epsilon}|^2 L_0(E_s + E_\epsilon) \vert s \rangle\vert \epsilon\rangle\langle\epsilon\vert\langle s\vert, \end{equation}

\noindent and $\hat \rho_{L}^{0,\mathrm{coh}}$ gives the coherences between the $\vert s \rangle \vert \epsilon\rangle$ states with trace of zero,

\begin{equation}  \hat \rho_{L}^{0,\mathrm{coh}} =  \sum_{s,\epsilon \neq s', \epsilon'}\tilde g_{s,\epsilon} \tilde g_{s',\epsilon'}\sqrt{L_0(E_s + E_\epsilon) L_0(E_{s'} + E_{\epsilon'}) } \vert s \rangle\vert \epsilon\rangle\langle\epsilon'\vert\langle s'\vert .\end{equation}

First consider the diagonal component $\hat \rho_{L}^{0,\mathrm{diag}}$. Its time average is

\begin{equation} \label{rho diag 1} \langle \hat \rho_{L}^{0,\mathrm{diag}} \rangle_{t\to\infty} = \sum_{s,\epsilon}  |\tilde g_{s,\epsilon}|^2 L_0(E_s + E_\epsilon) \langle \hat \rho_{s,\epsilon}\rangle_{t\to\infty}, \end{equation}

\noindent where $ \langle \hat \rho_{s,\epsilon}\rangle_{t\to\infty}$ is the time-average of a single $\vert s \rangle\vert \epsilon\rangle$ initial state.  Using the result for $ \langle \hat \rho_{s,\epsilon}\rangle_{t\to\infty}$ from Eq.~\ref{se eq}, then approximating the sum as a convolution integral analogous to Eq.~\ref{convolution integral} this gives

\begin{equation} \langle s' \vert \langle \epsilon' \vert \langle \hat \rho_{L}^{0,\mathrm{diag}} \rangle_{t\to\infty} \vert \epsilon'\rangle \vert s'\rangle \approx L_f(E_{s'} + E_{\epsilon'}), \end{equation}

\noindent where $L_f(E_{s'} + E_{\epsilon'})$ is the final Lorentzian in Eq.~\ref{Psi eq Lorentzian}.  Thus, we have arrived at the final Lorentzian distribution by considering the time-averaged diagonal component of the density operator.  We now consider the coherence component of the density operator in Eq.~\ref{Lorentzian density op initial}. 

The coherence component $\langle \hat \rho_{L}^{0,\mathrm{coh}} \rangle_{t\to\infty}$ of the time-averaged density operator has trace zero, so it has no contribution to the total probability of the time-averaged state and only serves to give fluctuations about the diagonal component $\langle\hat \hat \rho_{L}^{0,\mathrm{diag}}\rangle_{t\to\infty}$, with zero average fluctuation per basis state.  Based on the average behavior, we will simply approximate time-average of the coherence term as zero

\begin{equation}  \langle s' \vert \langle \epsilon' \vert\langle\hat \rho_{L}^{0,\mathrm{coh}}\rangle_{t\to\infty} \vert \epsilon'\rangle\vert s'\rangle \approx 0 .\end{equation}

\noindent  We will find that this approximation works very well to model our results.  Similarly, Deutsch treated $\langle \hat \rho_{L}^{0,\mathrm{coh}}\rangle_{t\to\infty}$ as negligible when calculating operator expectation values, in Eq.~5.7 of Ref.~\cite{DeutschSupplemental}.  

From the analysis of this section, the average equilibrium distribution for the initial Lorentzian state of Eq.~\ref{Psi0 Lorentzian} is  

\begin{equation} \label{Lorentzian eq}  \langle s' \vert \langle \epsilon' \vert\langle \hat \rho_{L}^0 \rangle_{t\to\infty}\vert \epsilon'\rangle\vert s'\rangle \approx L_f(E_{s'} + E_{\epsilon'}). \end{equation}

\noindent This gives the final average Lorentzian in the time-evolved state of Eq.~\ref{Psi eq Lorentzian} and Fig.~\ref{Lorentzian state fig}.  We will now consider the fluctuations about the Lorentzian average, to devise the fluctuation terms  $\tilde g_{s,\epsilon}$ in the final expression for equilibrium Lorentzian state of Eq.~\ref{Psi eq Lorentzian}.

\subsubsection{Fluctuations  in the coefficients of the time-evolved Lorentzian state} \label{lllllll}
\stepcounter{section}
In this section we analyze the fluctuation terms $\tilde g_{s,\epsilon}$ in the expression for the final Lorentzian state of Eq.~\ref{Psi eq Lorentzian}.  
Following the same reasoning as in Section \ref{se state fluctuation section}, we assume that the fluctuations $\tilde g_{s,\epsilon}$ can be expressed in the form of Eq.~\ref{g decomposition}.  We analyze the real and imaginary components $g_{s,\epsilon}$ and $g_{s,\epsilon}'$ using relations analogous using Eqs.~\ref{re g} and \ref{im g} but with the final state Lorentzian of Eq.~\ref{L0 final},

\begin{equation} \label{re g 2} g_{s,\epsilon} = \mathrm{Re}\left( \frac{ \tilde g_{s,\epsilon}}{1/\sqrt{2}}\right) = \mathrm{Re} \left(\frac{c_{s,\epsilon}(t)}{\sqrt{L_f(E_{s} + E_{\epsilon}))/2}}\right) \end{equation}

\noindent and

\begin{equation} \label{im g 2} g_{s,\epsilon}' =  \mathrm{Im}\left( \frac{ \tilde g_{s,\epsilon}}{1/\sqrt{2}}\right)  = \mathrm{Im}\left(\frac{c_{s,\epsilon}(t)}{\sqrt{L_f(E_{s} + E_{\epsilon})}}\right), \end{equation}

\noindent where $c_{s,\epsilon}(t)$ are the exact time-dependent coefficients of the $\vert s \rangle \vert \epsilon\rangle$ basis states taken at an instant in time $t$ at equilibrium (the results are similar for other $t$ at equilibrium).

\begin{figure}[h]
\begin{center}
\includegraphics[width=8cm,height=11cm,keepaspectratio,angle=0,]{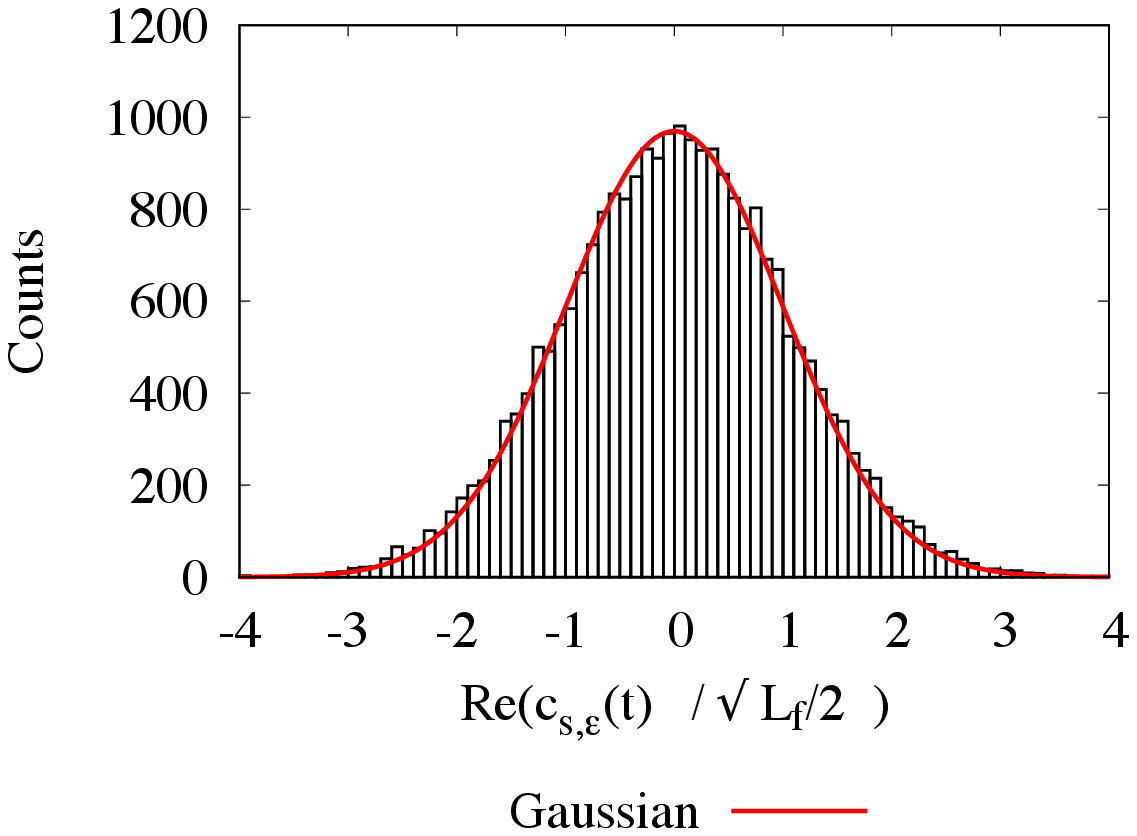}
\includegraphics[width=8cm,height=11cm,keepaspectratio,angle=0,trim={0 0 0 0},clip]{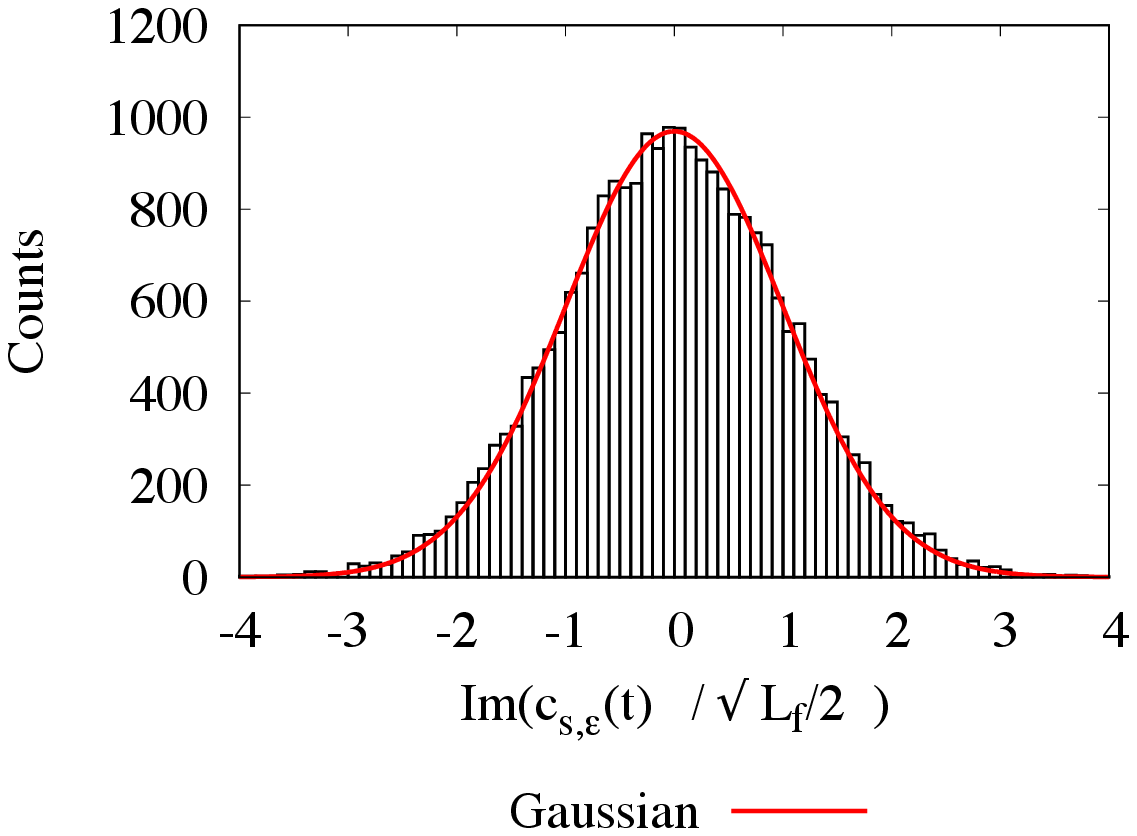}
\end{center}
\caption[Coefficient variations for a time-evolved Lorentzian initial state.]{Histogram counts of coefficient variations for a time-evolved Lorentzian initial state.  The real and imaginary parts the variations in Eqs.~\ref{re g 2} and \ref{im g 2} each follow a Gaussian distribution.}
\label{Lorentzian state coeff fig}
\end{figure}

Fig.~\ref{Lorentzian state coeff fig} shows the distribution of the $g_{s,\epsilon}$ and $g_{s,\epsilon}'$, taken as the right hand sides of Eqs.~\ref{re g 2} and \ref{im g 2}.  The distributions follow the standard Gaussian distribution, indicating that both $g_{s,\epsilon}$ and $g_{s,\epsilon}'$ behave as standard Gaussian variates, analogous to what we saw with the coefficients of the time-evolved $\vert s \rangle \vert \epsilon\rangle$ state in Section \ref{se state fluctuation section}.  

In total, we have seen in this section how the time-evolution of an initial Lorentzian state of Eq.~\ref{Psi0 Lorentzian} gives the final Lorentzian state of Eq.~\ref{Psi eq Lorentzian}, with random complex Gaussian variate fluctuations $\tilde g_{s,\epsilon}$ about the final Lorentzian $L_f$. 

\section*{C: Entropy of the Lorentzian} \label{Lorentzian entropy appendix}

In this section we derive the entropy, Eq.~30 of the main text, for a state with random variations about a Lorentzian, as in the previous sections and in Eqs.~22 and 27 of the main text.

Each of the Lorentzian states has squared coefficients of the approximate form 

\begin{equation} \label{Lorentzian palpha} p_\alpha = |c_\alpha|^2 \approx |\tilde g_\alpha|^2 L_\alpha =  |\tilde g_\alpha|^2\frac{1}{\pi} \frac{\gamma/\rho}{\Delta E_\alpha^2 + \gamma^2}, \end{equation}

\noindent where $\Delta E_\alpha = E_0 - E_\alpha$ is the energy difference between the basis state $\vert \alpha\rangle = \vert s \rangle\vert \epsilon\rangle$ and the initial state energy $E_0$, $\tilde g_\alpha$ is a complex Gaussian variate as in Eqs.~24-25 of the main text, and $\gamma$ is the half-width at half-max of the Lorentzian.  Using these coefficient distributions we will calculate the entropy from Eq.~6 of the main text, repeated here as

\begin{equation} S_{univ}^Q = -\sum_\alpha p_\alpha \ln p_\alpha. \end{equation}

\noindent Using Eq.~\ref{Lorentzian palpha} the entropy is

\begin{equation} \label{Sapprox} S^Q_{univ} \approx -\sum_{\alpha} |\tilde g_{\alpha} |^2L_{\alpha}\ln\left( |\tilde g_{\alpha} |^2L_{\alpha}\right) =  -\sum_{\alpha} |g_{\alpha} |^2L_{\alpha}\ln\left( L_{\alpha}\right) - \sum_{\alpha}L_{\alpha} |g_{\alpha} |^2\ln\left( |g_{\alpha} |^2\right)   \end{equation} 

\noindent The $g_{\alpha}$ are statistically independent from the $L_{\alpha}$ by assumption.  This suggests replacing the individual $|g_{\alpha}|^2$ in the first sum on the right of Eq.~\ref{Sapprox} with their average value $\langle |g_{\alpha} |^2\rangle = 1,$

\begin{equation} \sum_{\alpha} |g_{\alpha} |^2L_{\alpha}\ln\left( L_{\alpha}\right) \approx \sum_{\alpha} L_{\alpha}\ln\left( L_{\alpha}\right) , \end{equation} 

\noindent leaving just the entropy of the perfect Lorentzian.  For the second sum on the right of Eq.~\ref{Sapprox} the statistical independence of the $g_{\alpha}$ suggests replacing the $ |g_{\alpha} |^2\ln\left( |g_{\alpha} |^2\right) $ with the average value

\begin{equation} \sum_{\alpha}L_{\alpha} |g_{\alpha} |^2\ln\left( |g_{\alpha} |^2\right) \approx \langle  |g_{\alpha} |^2\ln\left( |g_{\alpha} |^2\right) \rangle \sum_{\alpha} L_{\alpha} =  \langle  |g_{\alpha} |^2\ln\left( |g_{\alpha} |^2\right) \rangle \end{equation}

\noindent where the last equality uses the normalization of the Lorentzian $ \sum_{\alpha} L_{\alpha} = 1$.  In total, Eq.~\ref{Sapprox} is then approximated as

\begin{equation} \label{Sapprox final} S^Q_{univ} \approx -\sum_{\alpha} L_{\alpha}\ln\left( L_{\alpha}\right)  -  \langle  |g_{\alpha} |^2\ln\left( |g_{\alpha} |^2\right) \rangle \end{equation}

\noindent The first term is the entropy of a Lorentzian, while the second term gives the deviation from the perfect Lorentzian entropy due to the random variations in the state.

Now we will evaluate the terms in Eq.~\ref{Sapprox final}.  The Lorentzian sum in the first term can be approximated as the integral

\begin{equation} \label{Lorentzian entropy integral} -\sum_{\alpha} L_{\alpha}\ln\left( L_{\alpha}\right) \approx -\int_{-\infty}^{\infty} d(\Delta E_\alpha) L_{\alpha} (\Delta E_\alpha)\rho(E_{0}) \ln\left( L_{\alpha}(\Delta E_\alpha) \right) \end{equation}

\noindent In the integral approximation, the density of states $\rho$ is factored into the integrand to account for having approximately $\rho d(\Delta E_\alpha)$ states in the sum that are within each differential interval $d(\Delta E_\alpha)$ of integration.   To simplify the integral, we have approximated the density of states as the constant value at the central energy of the Lorentzian $\rho = \rho(E_{0})$, where the majority of probability in the Lorentzian is located.  

To evaluate the integral Eq.~\ref{Lorentzian entropy integral} we first split it into two separate integrals by factoring $\rho(E_0)/\rho(E_0)$ into the logarithm then separating out a term $-\ln \rho(E_0)$,

$$ \label{Lorentzian entropy integral 2} -\int_{-\infty}^{\infty} d(\Delta E_\alpha) L_{\alpha} (\Delta E_\alpha)\rho(E_0) \ln\left( L_{\alpha}(\Delta E_\alpha) \right)$$
\begin{equation}  = -\int_{-\infty}^{\infty} d(\Delta E_\alpha) L_{\alpha} (\Delta E_\alpha)\rho(E_0) \ln\left( L_{\alpha}(\Delta E_\alpha) \rho(E_0)\right) +  \int_{-\infty}^{\infty} d(\Delta E_\alpha) L_{\alpha} \rho(E_0) \ln \rho(E_0) \end{equation}

\noindent The first integral on the right of Eq.~\ref{Lorentzian entropy integral 2} has the well known solution $\ln (4\pi \gamma)$ for the entropy of a continuous Lorentzian distribution, while the second integral is simply $\ln \rho(E_0)$, since $\rho(E_0)$ is a constant and $\int_{-\infty}^{\infty} d(\Delta E_\alpha) \rho(E_0)L_{\alpha} (\Delta E_\alpha) = 1$ by the normalization of the Lorentzian.  Then in total we have

\begin{equation} \label{lorentzian entropy appendix} -\sum_{\alpha} L_{\alpha}\ln\left( L_{\alpha}\right) \approx  \ln (4\pi \gamma\rho(E_0)) \end{equation} 

The final term in Eq.~\ref{Sapprox final} for the average $\langle  |g_{\alpha} |^2\ln\left( |g_{\alpha} |^2\right) \rangle$ is calculated through integration over all the values of $g'$ and $g''$ with the Gaussian variate probability density $p(g) = (2\pi)^{-1/2} \exp(-g^2/2)$,

\begin{equation} \label{variation entropy appendix} \langle  |g_{\alpha} |^2\ln\left( |g_{\alpha} |^2\right) \rangle = \int_{-\infty}^\infty dg' \int_{-\infty}^{\infty} dg'' p(g')p(g'') \frac{g'^2 + g''^2}{2} \ln \frac{g'^2 + g''^2}{2} = g_0  \end{equation}

\noindent where 

\begin{equation} g_0 = 1-\gamma_{EM} \end{equation}

\noindent where $\gamma_{EM}= 0.577\ 215...$ is the Euler-Mascheroni constant.  

Putting Eqs.~\ref{lorentzian entropy appendix} and \ref{variation entropy appendix} into Eq.~\ref{Sapprox final} gives the entropy for the Lorentzian states, in Eq.~30 of the main text,

\begin{equation} S \approx \ln (4\pi \gamma\rho) - g_0 . \end{equation}

\bibliography{main.bib}
\bibliographystyle{unsrt}

\end{document}